# A Line is Not Just a Line: Visualizing Harmonious Complex Line Sets

**Francesca Samsel**
Texas advanced Computing Center
University of Texas at Austin, USA
fsamsel@tacc.utexas.edu

**Daniel Keefe**
University of Minnesota, USA
dfk@umn.edu

**Gregory Abram**
Texas advanced Computing Center
University of Texas at Austin, USA
gda@tacc.utexas.edu

## Abstract

Line is a foundational building block in visual design for traditional media and computer graphics. Controlled variation in line quality can be used to define form, create abstractions, distinguish categories, control affect, and more. Yet, when it comes to computer graphics visualizations of science, particularly those built upon 3D computer graphics, too often a line is just a line. This portfolio presents a series of visual design studies and lessons learned that explore the diverse character of line that might be useful for 3D visualizations. The imagery begins to show how more sophisticated use of this powerful visual building block could increase the clarity of data visualizations while also bringing a human, handcrafted aesthetic that reinforces the people and impacts behind scientific climate data.

## Introduction

From visualizing streamlines, streaklines, and pathlines in fluid flows to neural pathways in the brain, lines take center stage in many computer-generated data visualizations–but not with the aesthetic explored in this pictorial!

Our focus is on global climate data, the type that scientists study today with cutting-edge computer models. It should be clear that every living organism has a stake in these data and the new knowledge they help produce. We might, therefore, expect visuals used to study, discuss, and disseminate the science to reinforce our high stakes, collective ownership, and care. Unfortunately, typical practice is just the opposite as data represented with acidic palettes or sterile, geometric forms. There is a standard justification for this aesthetic choice–we must avoid "biasing" the data. As members of the VIS scientific research community, we are well trained and understand the potential for bias and rationale for this statement, but in the context of today's urgent need for collective care, our response is twofold. First, all data are biased, and the best way to overcome these biases is to reveal and discuss them. Second, our most pressing current problem is that we are not engaging in these discussions, and one reason may, in fact, be that the way we visualize science makes it appear untouchable, other-worldly, automatic, and algorithmic.

Too often our visualizations remove any evidence of our relatives and relationships (human and other-than-human).

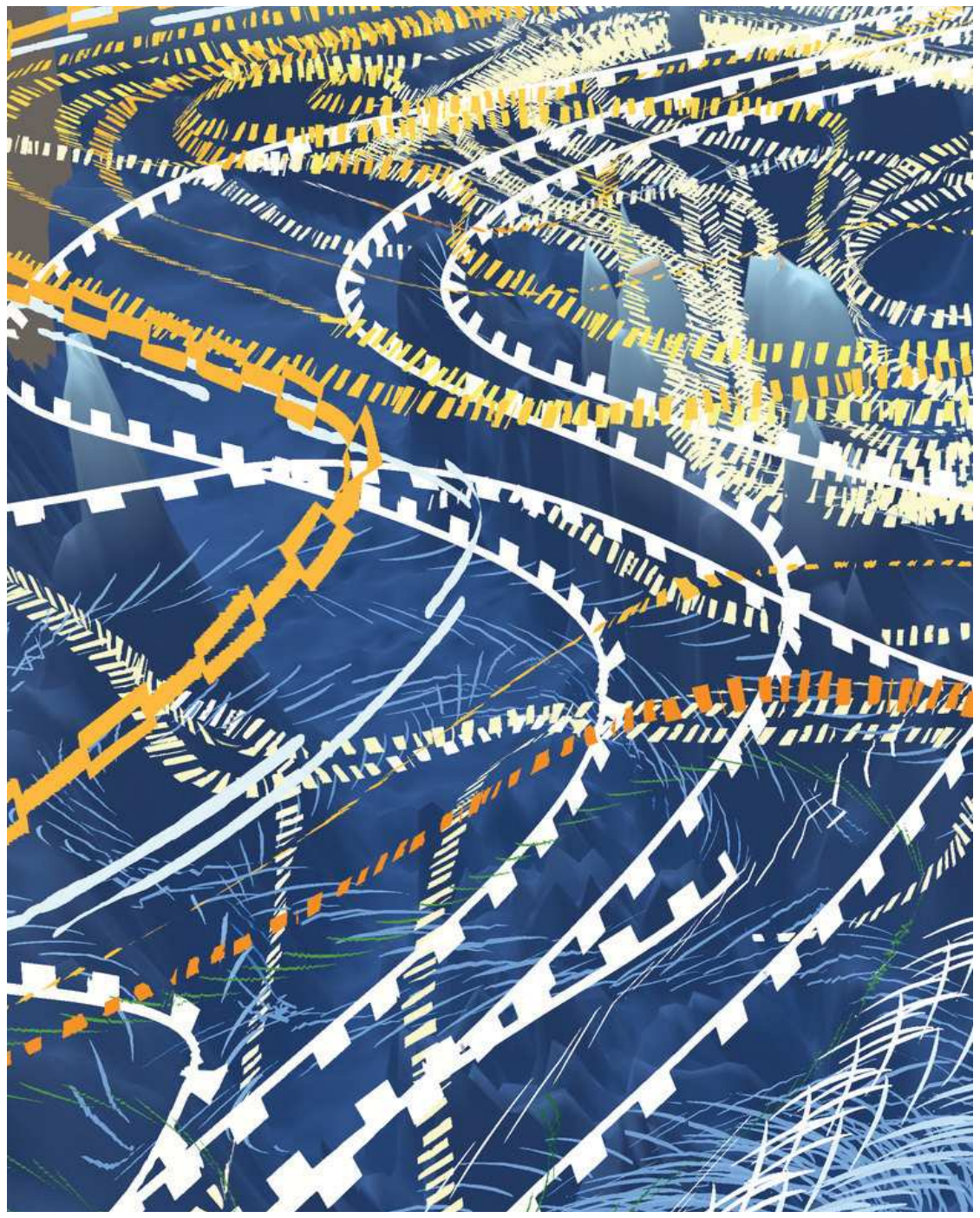

Figure 1: Five currents within the Gulf of Mexico.

## Section One: Background and Motivation

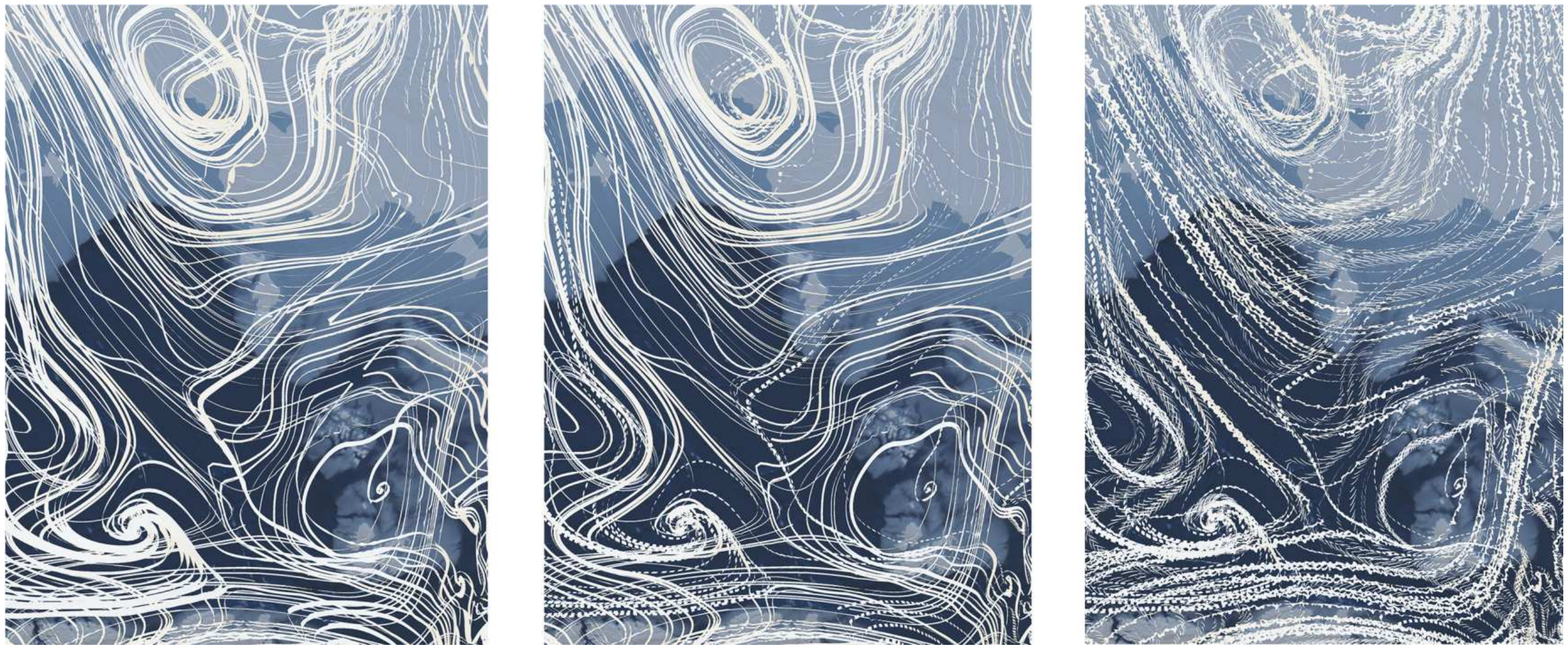

Figure 2: Visualization of three elevations of wind currents over the Arctic. The progression moves from standard lines (left), to a mix of artistic and standard lines (middle), and all custom line textures (right).

Below, it continues with a detailed view of the same data in with the lines colored in muted tones.

This portfolio documents an alternative approach that is decidedly natural and relatable. We present a deep visual investigation of the role of line in high-end, accurate, computer-generated, environmental science visualizations but with a new and expansive visual vocabulary of hand-drawn lines.
The data pictured in this portfolio come from partnerships with the U.S. Department of Energy climate scientists developing cutting-edge supercomputer models.

As we explore the potential of a radically different aesthetic, it is important to emphasize that the goal is not just “science communication” or dissemination of results. This work is intended to improve the capabilities of scientists to make discoveries. The data pictured come from partnerships with environmental scientists developing cutting-edge supercomputer models. The datasets are large and contain tens of data fields. Flow lines must be derived from the raw data by integrating through raw vector data files, like velocity or vorticity. The level of data processing involved is such that it is often quite a significant result to render a first picture of the data. However, our approach is guided by the principle that what scientists (and our planet) need is not just that first picture but one that has undergone a serious visual design refinement process that draws upon art and design theory and methods. The visualization objective scientists repeatedly emphasize is to see multiple variables in a single picture. Thus, with a focus on line, this raises visual design questions, such as: Can we design visually distinct sets of lines that have a more natural, recognizable aesthetic, and if so, what are the underlying contrast properties of those sets? What guidelines might designers use to harness the power of a more complex visual language without overloading viewers?

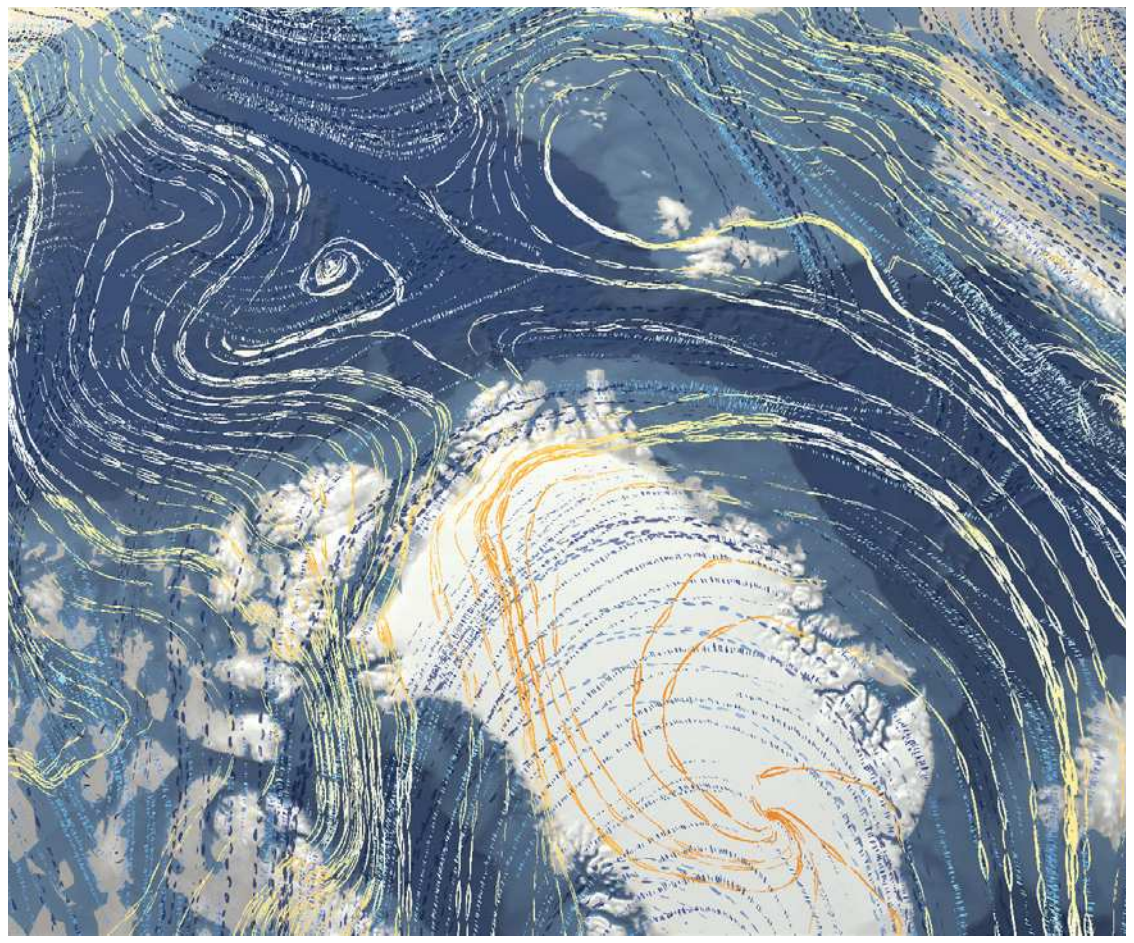

This portfolio documents examples and lessons learned from several years of deeply studying these questions. We begin in Section 2 with a Primer on Visual Contrast for Line Visualizations. Section 3 reports on our main findings organized as a series of visual studies, Visual Studies: Applying Contrast to Complex Data. These annotated examples demonstrate how to apply the foundational theory of Section 2, in context, in order to produce visualizations with clarity and harmony.

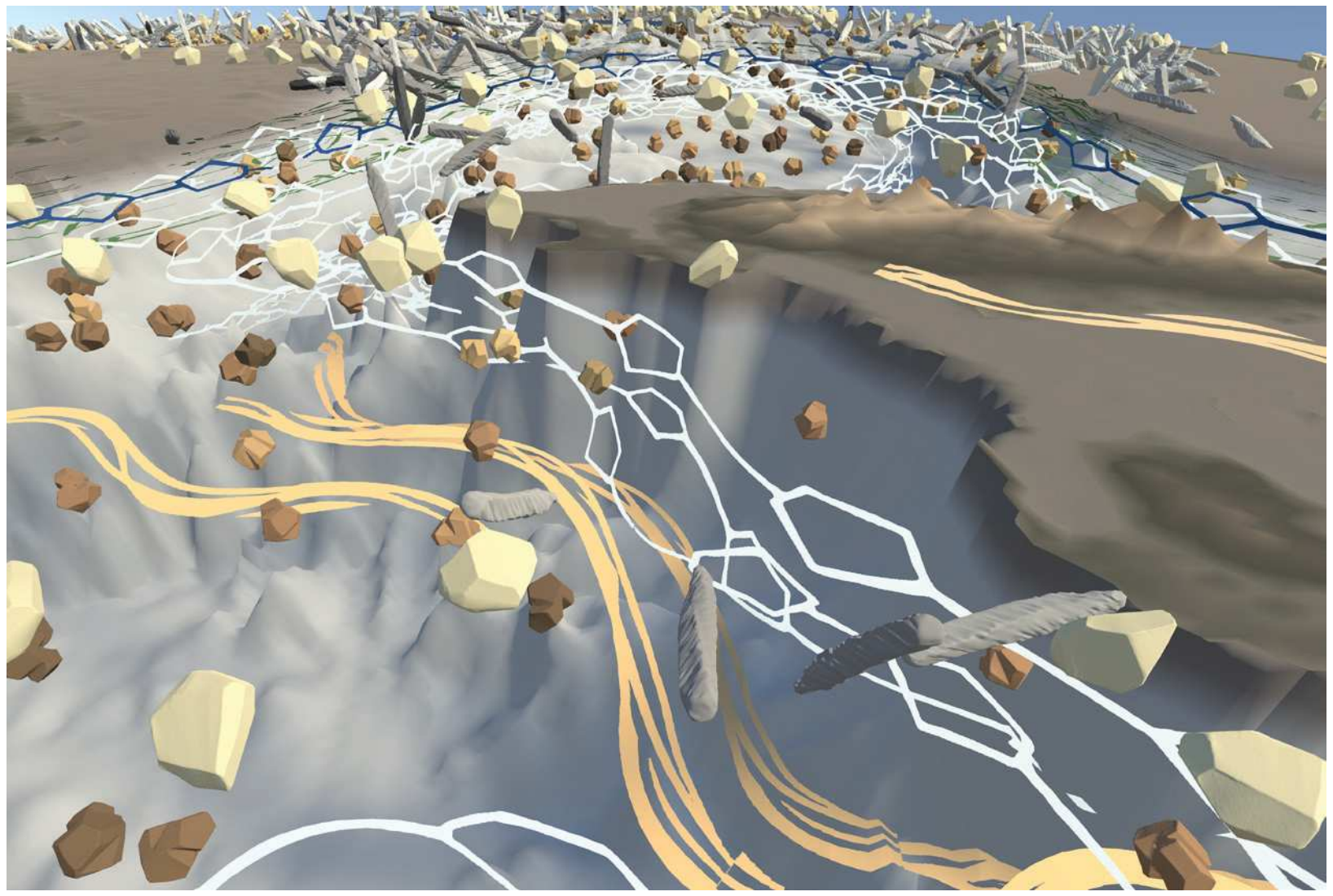

Figures 3 and 4: Artists know "a line is not just a line". As shown here in this pair of images, one change in the way a line is rendered can have a dramatic impact on the way we read a visualization.

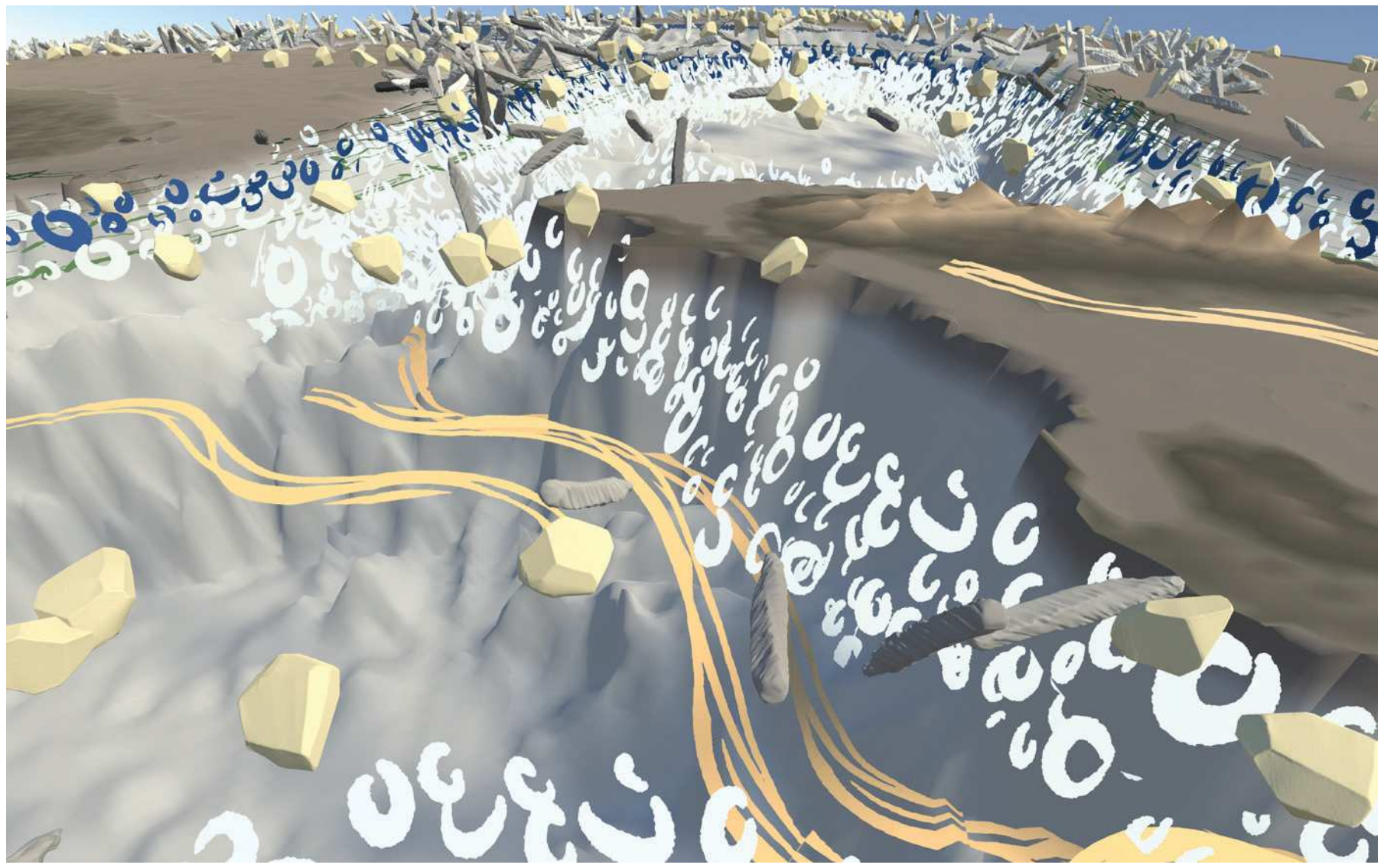

Now, before diving into the theory and examples, a final, brief note on our methodology. Our work is grounded in art practice and visual design while also enabled by new technologies. Specifically, using the Artifact-Based Rendering [1] framework, gives us the ability to define the visual building blocks of 3D data visualizations using traditional artistic media (e.g., ink, paint, clay, found objects), which opens up a much larger range of possibilities for exploring visual contrast in lines, as compared to traditional 3D visualization software, while also rendering the results with accurate, data-driven 3D graphics. The methodology includes scanning the 2D and 3D artifacts created by artists. Then, custom 3D computer graphics shaders place the digitized artifacts in space according to 3D data while also re-coloring, texturing, morphing, and adjusting other visual qualities in response to data variables. Artifact-Based Rendering supports a variety of line styles, including "oriented ribbons" (textured triangle strips) that follow 3D paths or even "dotted" lines that are composed by a series of sculpted clay glyphs arranged in space to follow a 3D path. The examples in this portfolio use oriented ribbons with data driving their placement, orientation, width, color, and texture. The powerful impact of the texture applied as a transparent cutout (i.e., alpha mask) on the 3D ribbon forms can be seen in these images.

## Section 2: Primer on Visual Contrast for Line Visualizations

The concept of contrast from Artistic Design Theory and Gestalt Theory [2, 3] is a useful way to explain (and guide the design of) many aspects of what makes an image work. For example, artists might describe how contrasting styles of line are used to clarify forms and lighting, imply movement, direct attention, signify importance, and/or convey emotion. Design theory identifies many types of contrast or ways that contrast can be achieved. In this portfolio, we explore the following six foundational types of contrasts in line, also illustrated in Figure 5:

- open – closed
- light – bold
- linear – curvilinear
- continuous – implied
- geometric – organic
- simple – complex

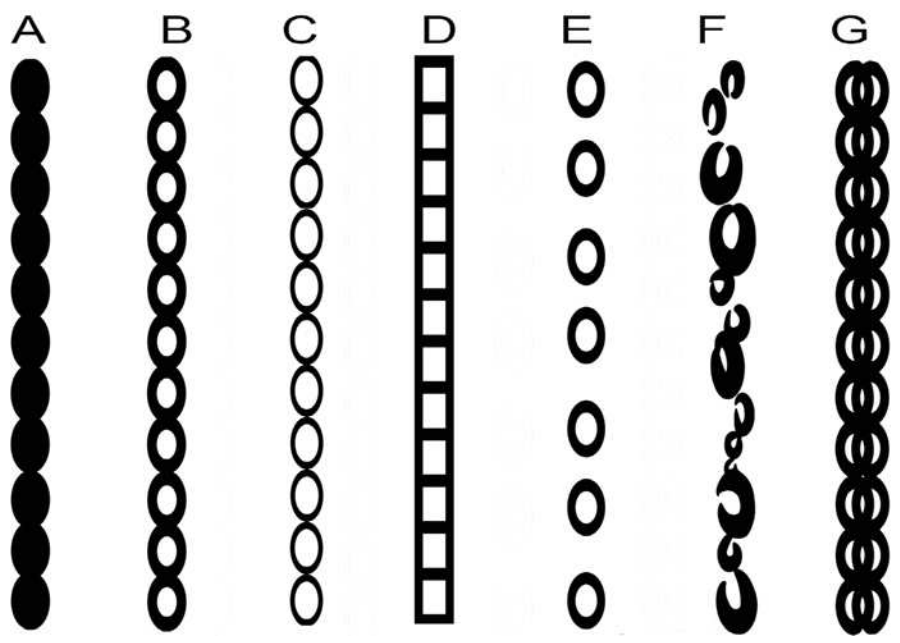

Figure 5: A can be described as: closed, bold, curvilinear, continuous, geometric, and simple. In contrast, B is open as opposed to closed; C is light instead of bold; D is linear instead of curvilinear; E is implied rather than continuous; F is organic rather than geometric; and, G is complex rather than simple.

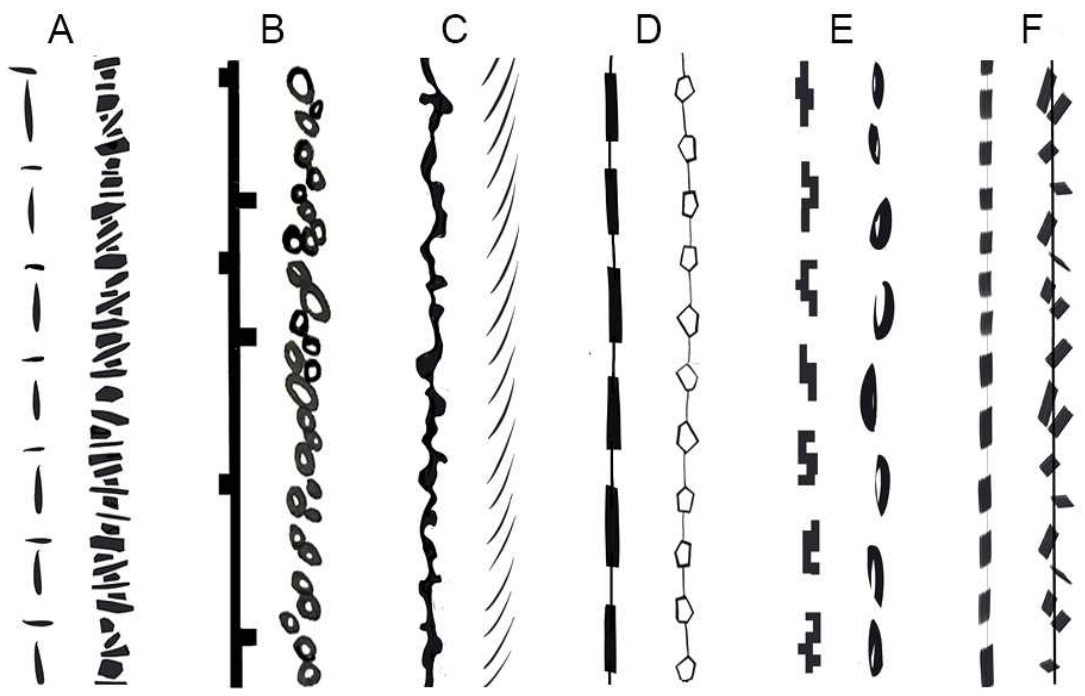

Figure 6: In practice, artists will naturally apply multiple types of contrast when designing lines with visualization in mind. Here, the primary type of contrast illustrated in each pair of lines is as follows. A: thin vs. thick; B: linear vs. curvilinear; C: continuous vs. implied, D: closed vs. open, E: geometric vs. organic; and F: simple vs. complex.

Already, these simple examples provide a richer palette of options for encoding data than what we typically see in 3D spatial visualization tools, which tend to use a standard simple, geometric, linear, continuous, closed line–perhaps with some control over the line width (i.e., light–bold contrast). However, where contrast really shines is when we consider variation across multiple types of contrast, as in Figure 6. When lines with contrasts like these are used in a data visualization, as in Figure 7, it is easy to visually separate the lines into categories, and this provides great opportunities for data encodings. More complex examples are shown in Figures 8 and 9.

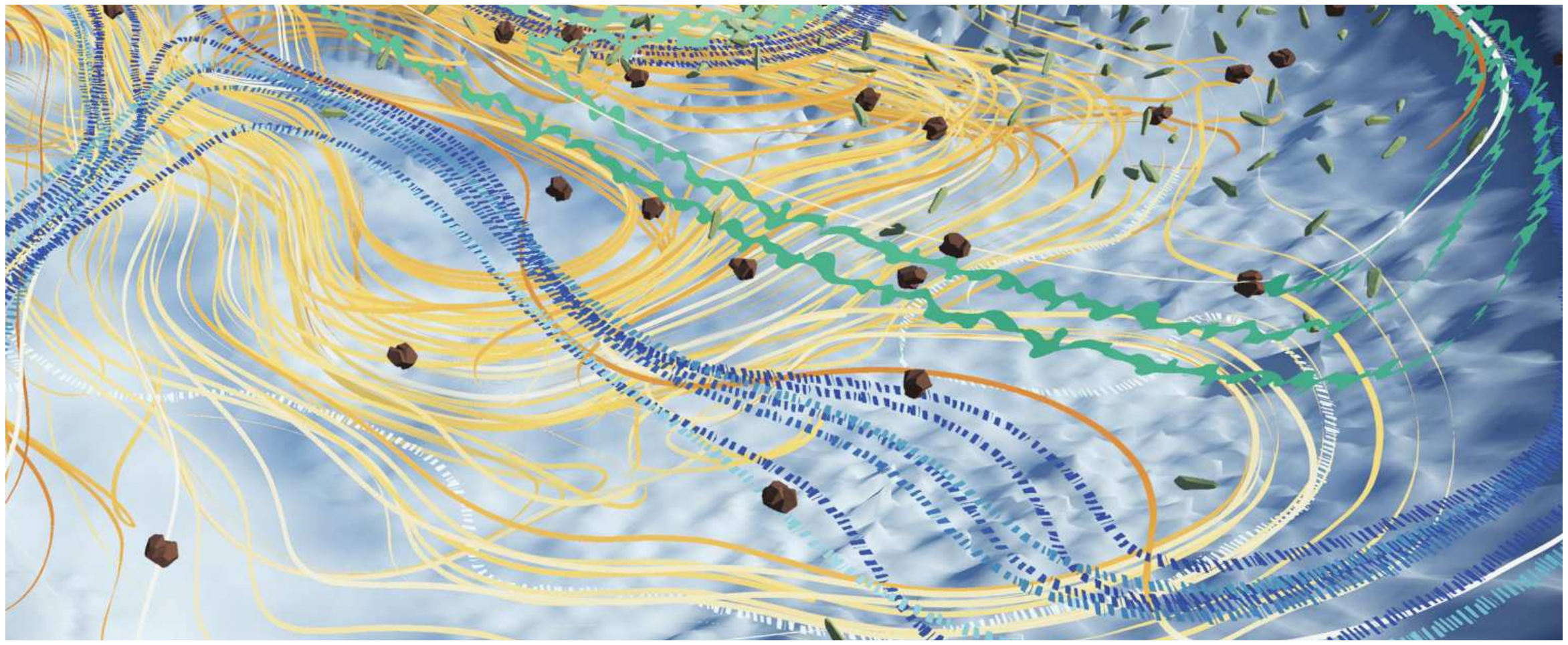

Figure 7: When visualizations are composed of lines that contrast according to these principles, it is possible to create visual differences that can encode data. For example, three distinct categories of lines are immediately visible in this visualization.

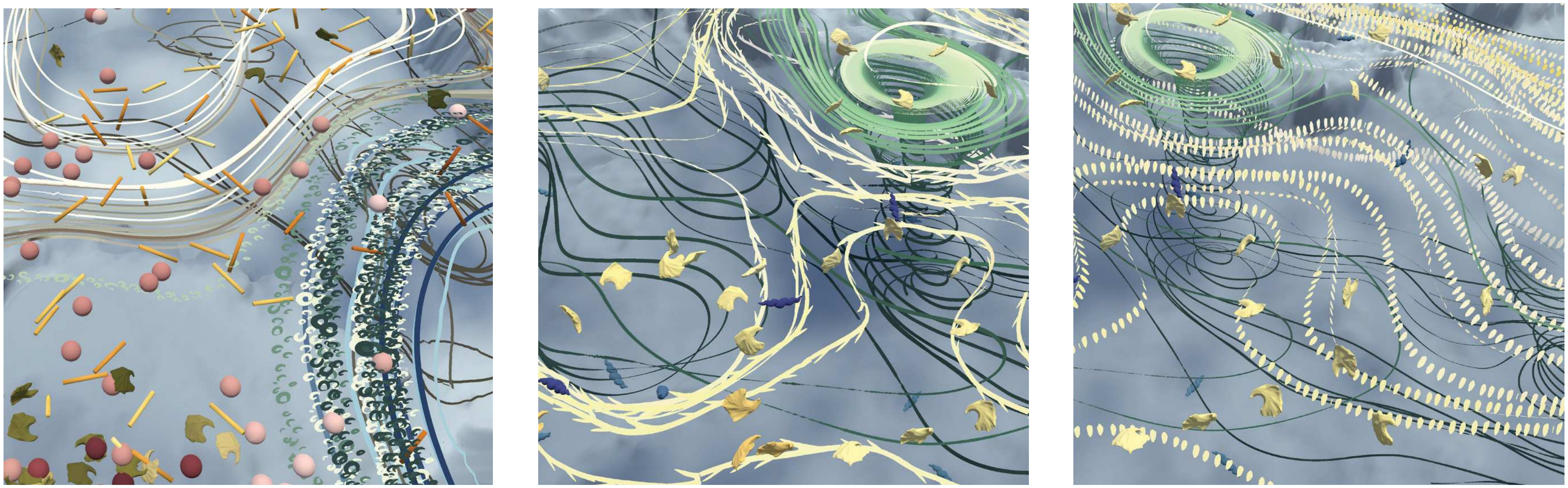

Figure 8: Variation across multiple types of contrast helps to distinguish lines. All three images include one set of rather standard (linear, continuous, geometric) lines as well as a second set of lines that contrast in style. The two sets are easily distinguished in all cases, but the types of contrast vary. In A, the second set is curvilinear, implied, organic, and complex. In B, the second set is curvilinear, organic, complex, and open. In C, the second set is curvilinear, organic, and complex.

Designing visualizations to incorporate a wider visual vocabulary of lines is a promising but challenging prospect. Formalizing the types of visual contrast helps to identify the design space while also emphasizing that this is a complex, multidimensional design space (e.g., a single line can be characterized as open, implied, geometric, and simple). Up this point contrasts have been described as binary choices (e.g., a line is either open or closed). Some contrasts, such as boldness, are continuous. Others contain both characteristics such as a line with both linear and curvilinear elements.

In general, lines will be more easily distinguished as contrast is increased and/or multiple types of contrast are introduced [3]. However, there are complex interactions when multiple types of contrast are applied across multiple lines, that we are just beginning to understand. See pages 12 and 13. An analogy is the way color works in visualization. Color can be characterized by its components (e.g., Hue, Saturation, and Value), but variation in these components does not tell the whole story of how color is perceived in an image–the components are not equally weighted in our interpretation of the difference between two colors, they interact with each other in complex ways, and our perception of color also always depends on the surrounding context. Thus, like color, lines can be a complex but super powerful tool for visual data encodings.

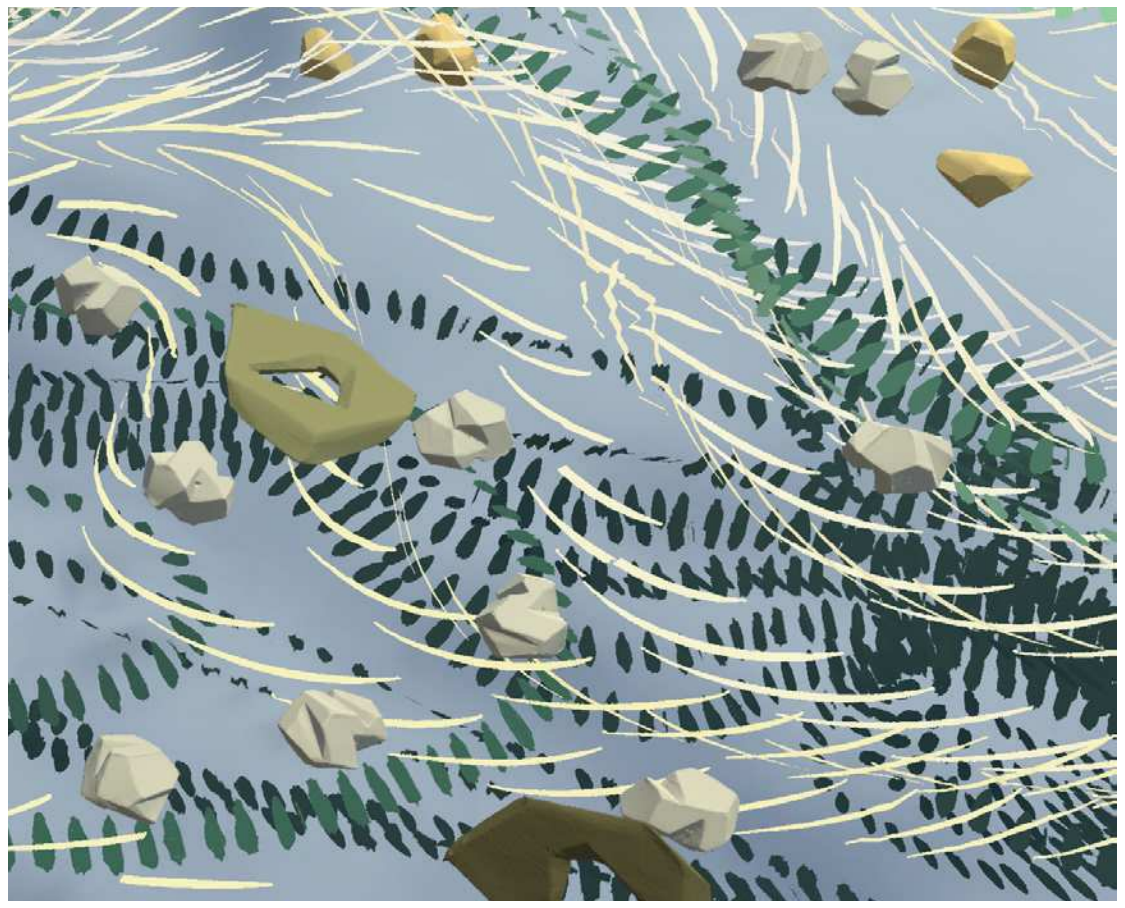

Figure 9: A more nuanced case. The two styles of line in this image are created from similar repeated marks. Both are implied, curvilinear, and organic. However, the changes in aspect ratio and scale lead to a light vs. bold contrast that is easily discernible.

## Section 3: Visual Studies: Applying Contrast to Complex Data

This page begins a series of examples best understood as visual studies to explore the potential use of a richer palette of lines to create accurate and engaging spatial data visualizations. Each study applies the types of contrast identified in Section 2 to real-world, actively studied global climate data.

### The Study of Line Weight

Line weight is often understood as the width of the line, but line weight also refers to the visual attention a line commands [3] as demonstrated in Figures 11 to 13. Understanding the impact of factors other than width is critical to directing attention, creating balance and or highlights within a visualization. For example, in Figure 10 the green line is equal in weight despite the larger width of the yellow line. In Figure 11, the green implied line, made of solid oval forms carries more weight than the yellow wispy curves.

Several other characteristic contribute to line weight. In Figure 12 the complexity of the profile and aggressive nature of the green line of creating its dominant position. In both Figures 12 and 13 the width and complexity provide weight. However, in Figure 13 the green line recedes because of the thinness of the line and the openness of the pattern. It is worth noting that the complexity does not overwhelm the other data visualized because the elements making up the line are thin and not densely distributed.

Profile complexity has a significant impact on line weight, which aligns with our findings on glyph design and visual weight. Interestingly, in Figure 13 the line complexity does not overwhelm the other data visualized. Note, while the interior of the line is complex, the profile is relatively stable.

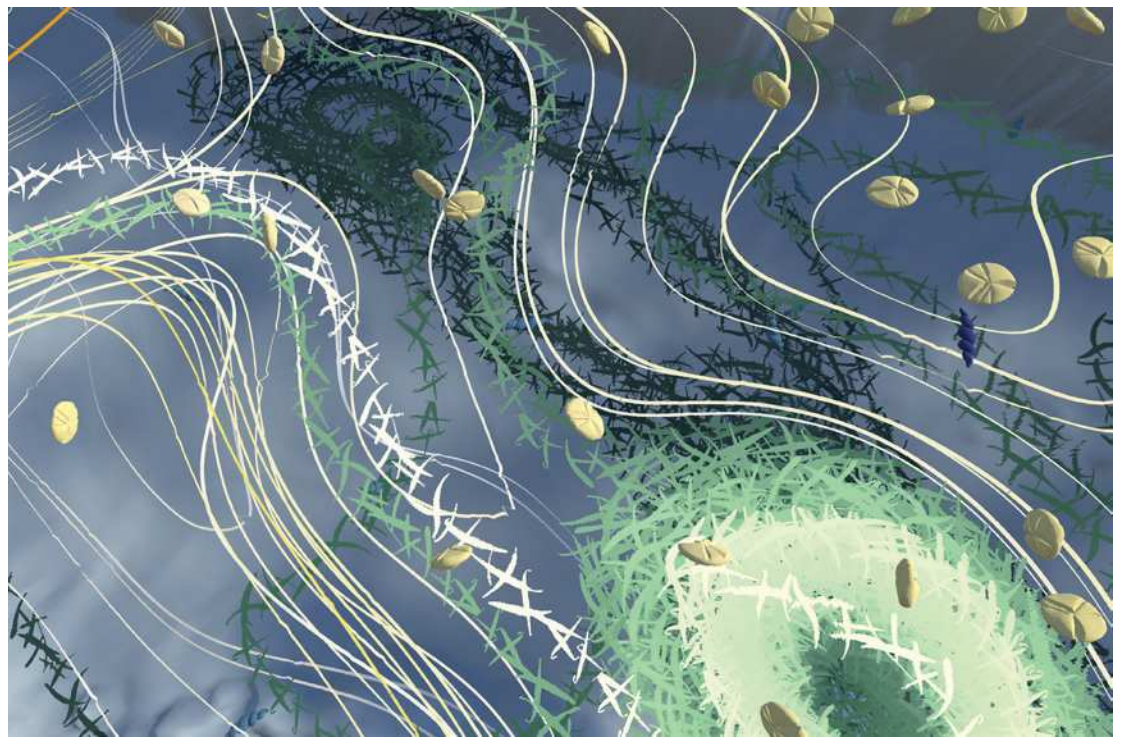

Figure 12: The aggressive affect of the green line along with density and complexity create impact.

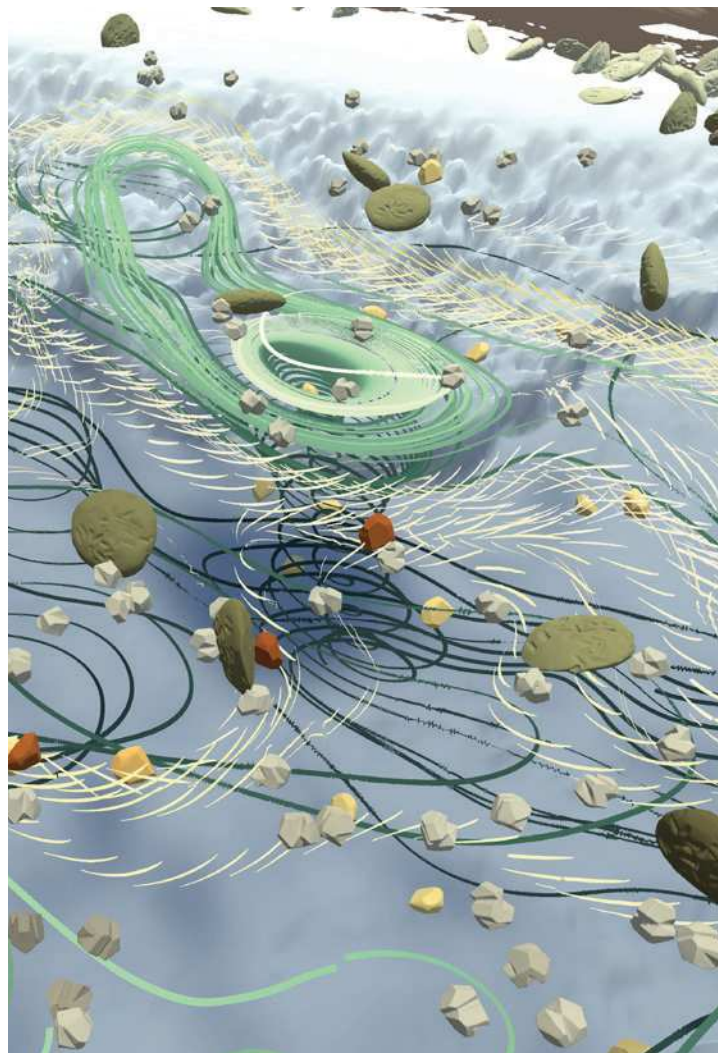

Figure 10: A standard line contrasts with open implied line

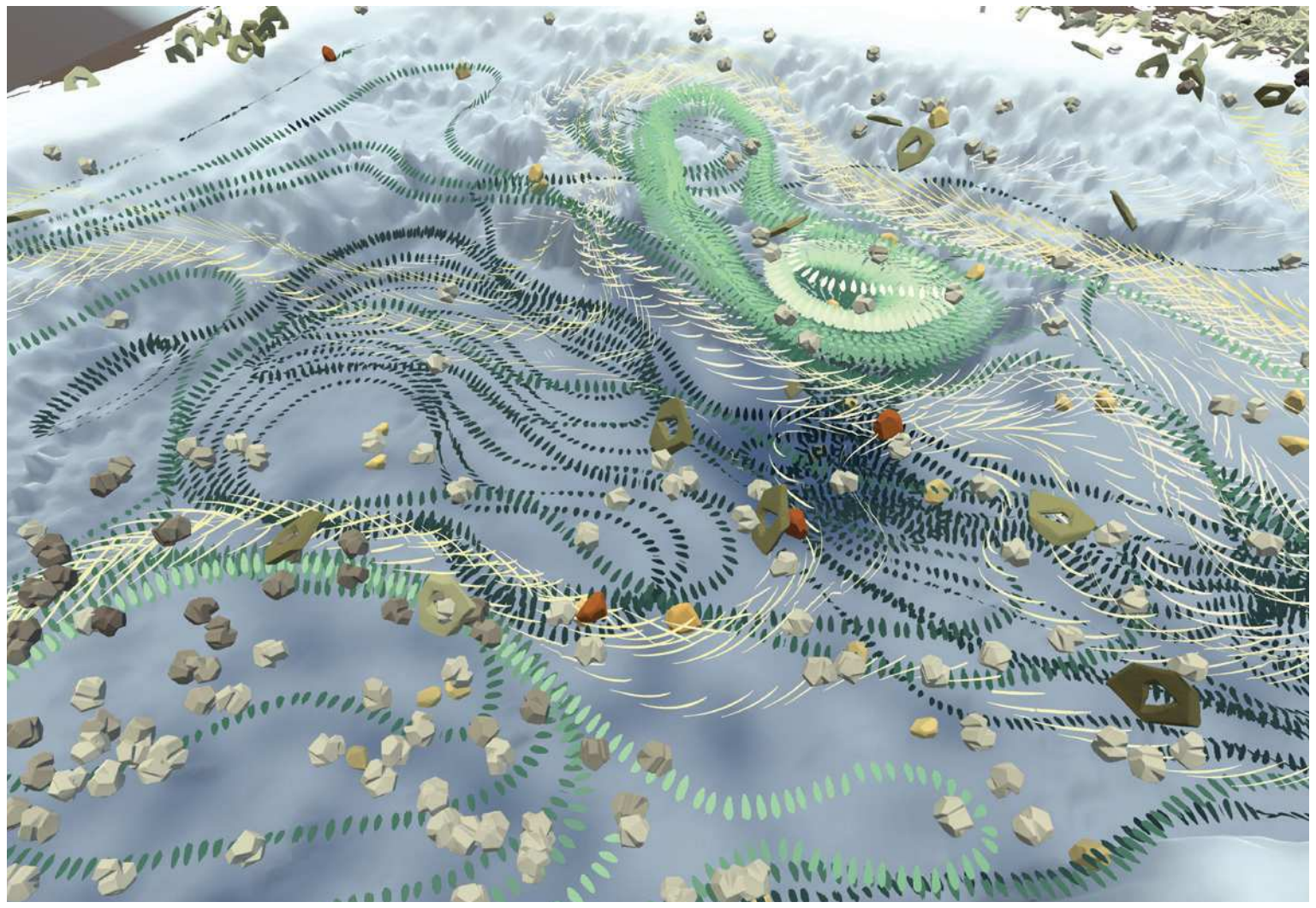

Figure 11: the elements that make up the green line are bolder, but not wider than the wispy yellow line implying motion.

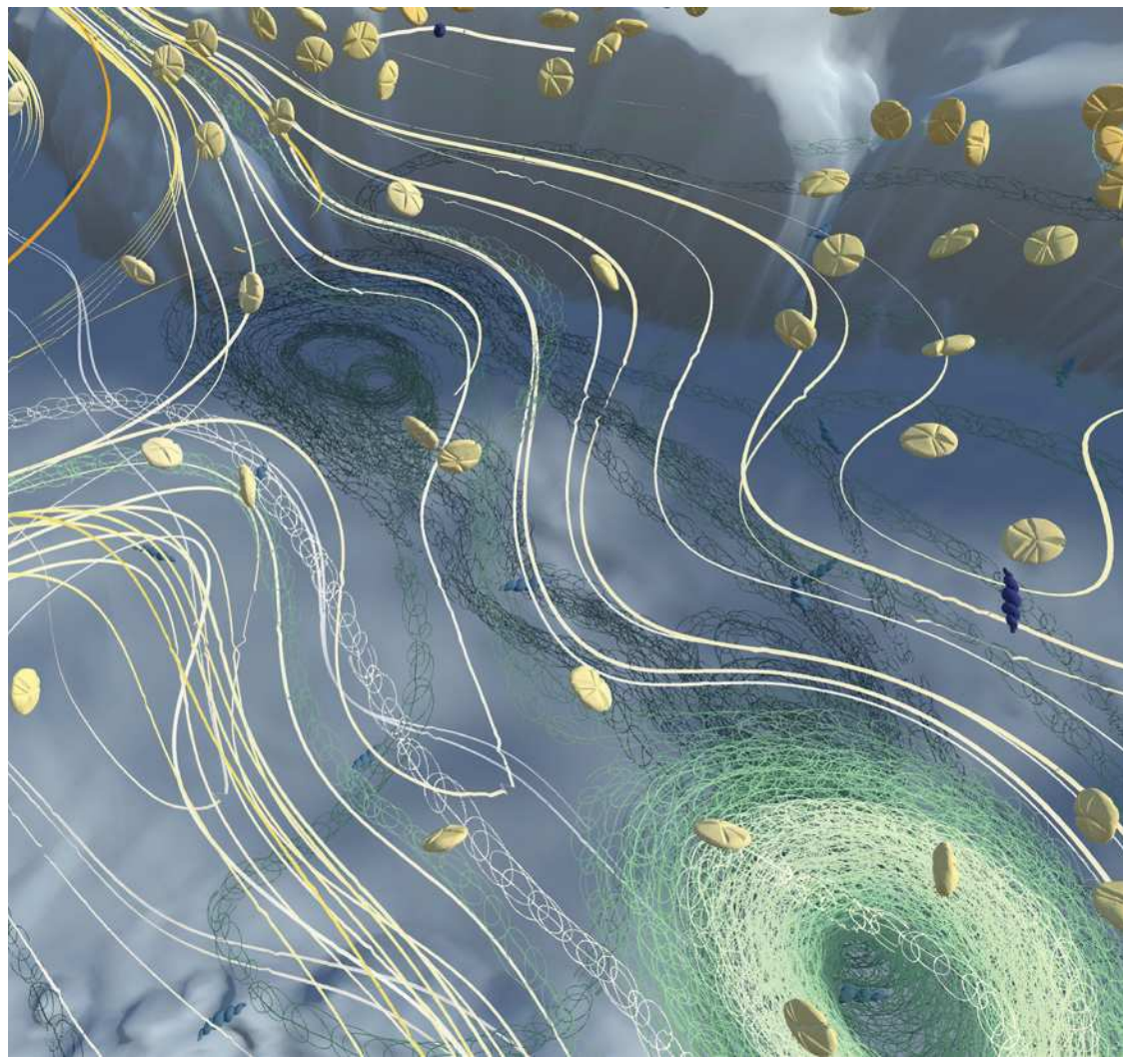

Figure 13: The circular pattern of the green line is calmer and thus quieter than in Figure 12.

## Linear Curvilinear, Organic Geometric, the Variety available within an Organic Palette

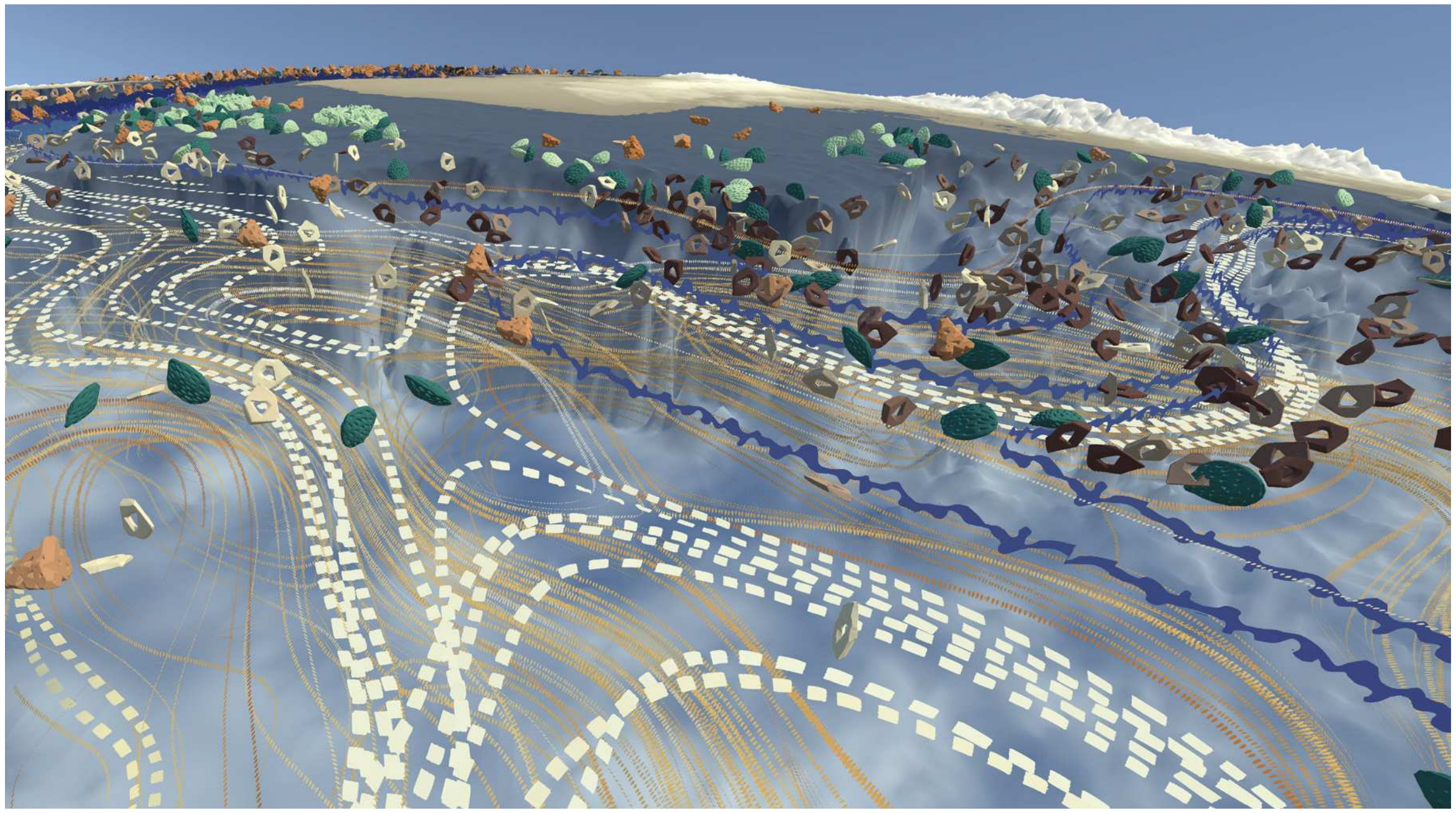

Figure 14:  The multiple types of contrast employed here: geometric verses organic, linear verses curvilinear and thick verses thin, create clear distinction between the three currents in this visualization.

Linear versus curvilinear is a well established potent type of contrast that is useful when one needs simple but strong contrast well documented in the literature [2, 3]. Another less-explored powerful contrast is geometric verses organic. Figures 8, 12 and 13 illustrate the power of this type of contrast.

Figure 16 contains glyphs that are geometric, organic and linear versus curvilinear. These clearly contrast with the associative lines in Figure 16.

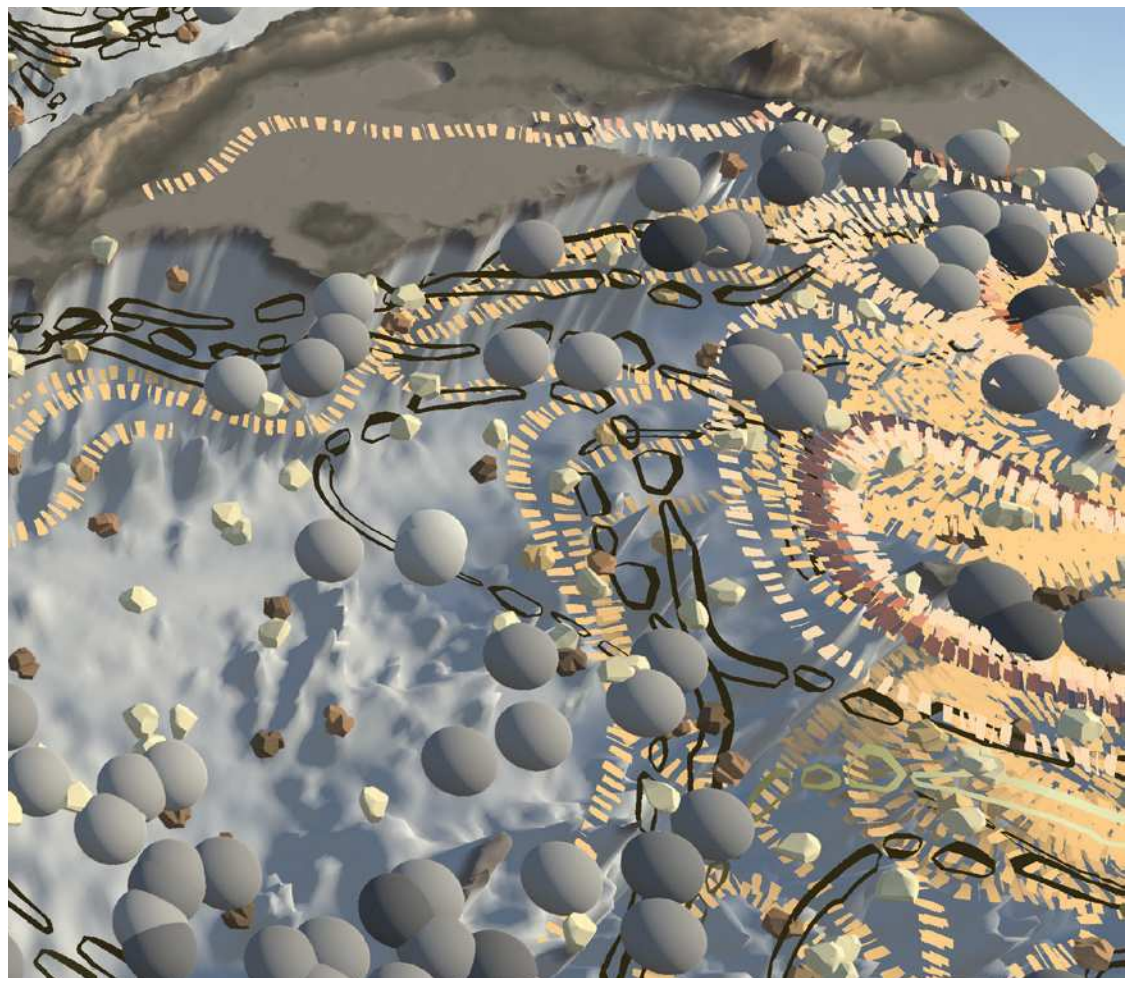

Figure 16: Geometric glyphs stand alone with in this very organic visualization.

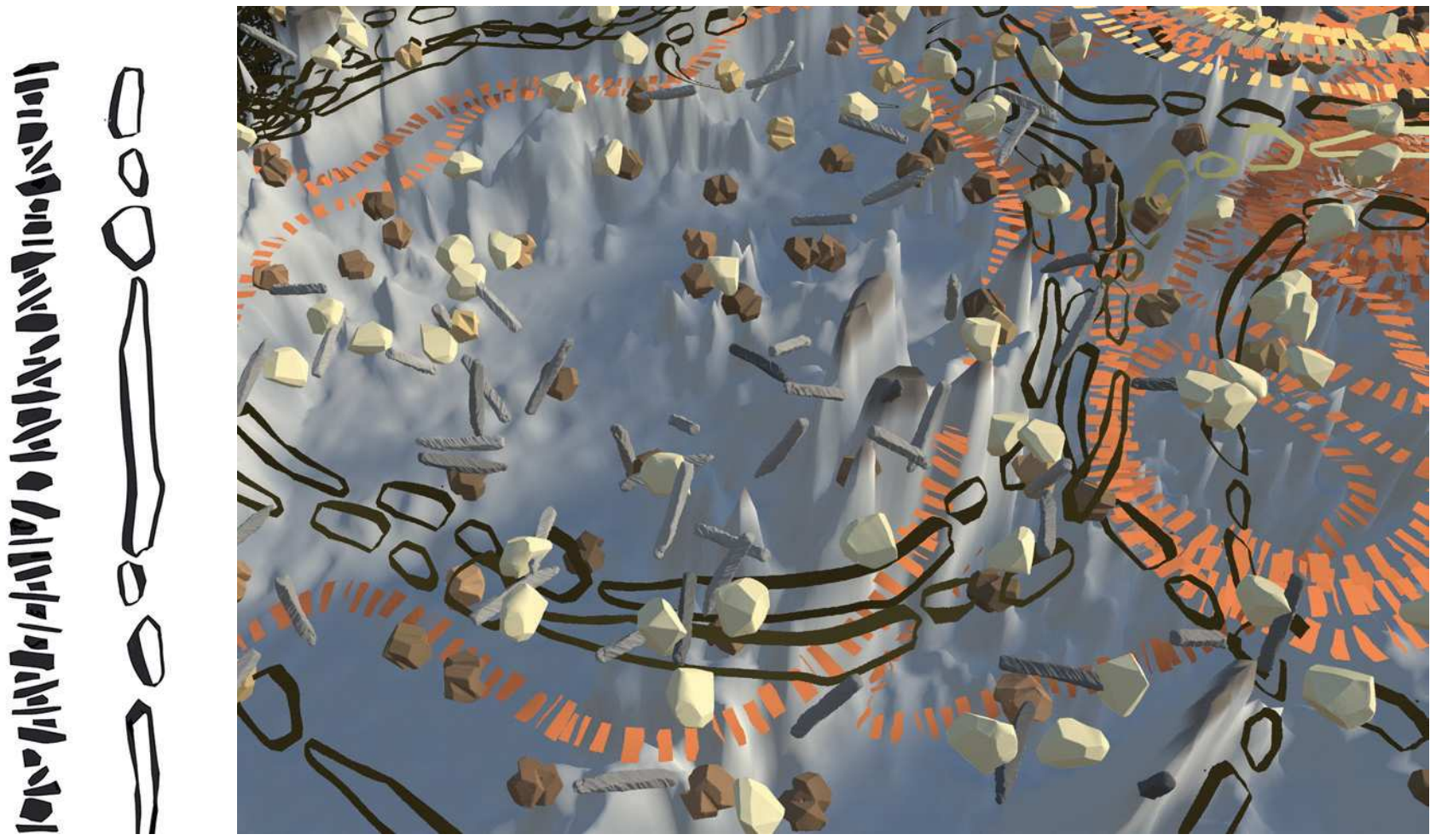

Figure 15 combines bold organic and bold geometric lines with lighter implied lines that add context.
The organic lines in Figure 14 reference rock formations and are complemented by glyphs that also suggest rocks. Organic forms are so rich and varied that one can find plenty of contrast within the organic alone.

## Study of Attention and Dominance

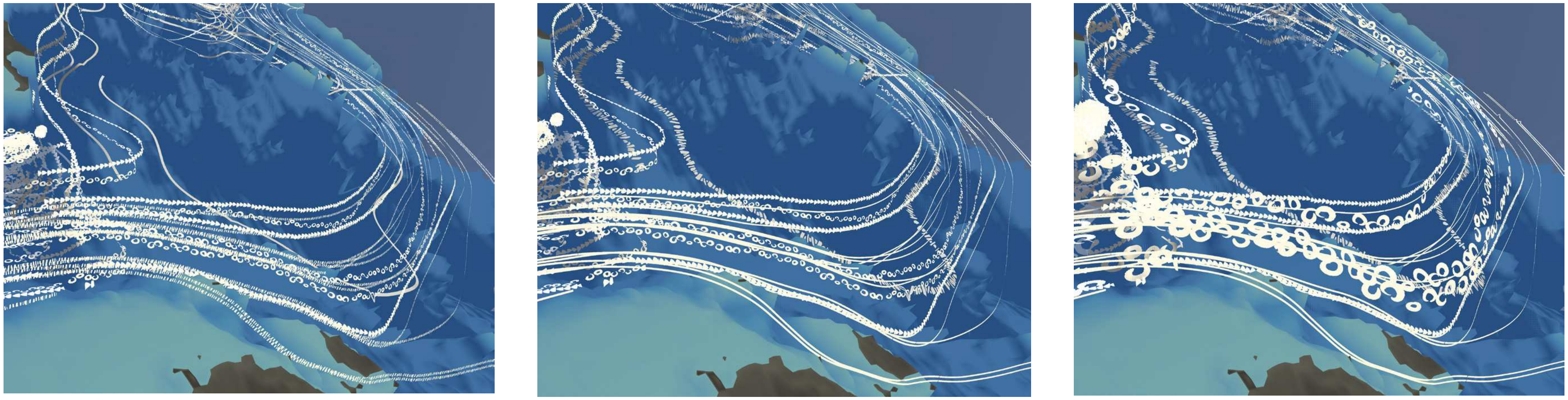

Figure 17: In comparing this set of three images, the first is the least successful for two reasons: First, the widths of the lines are too similar, and second, the many overlapping implied lines cause confusion. In the middle example, the implied, at the bottom of the image is exchanged with a plain line; this simple swap creates a clearer image. On the right, the width of the line implying water is significantly increased, creating a focal line and clarity.

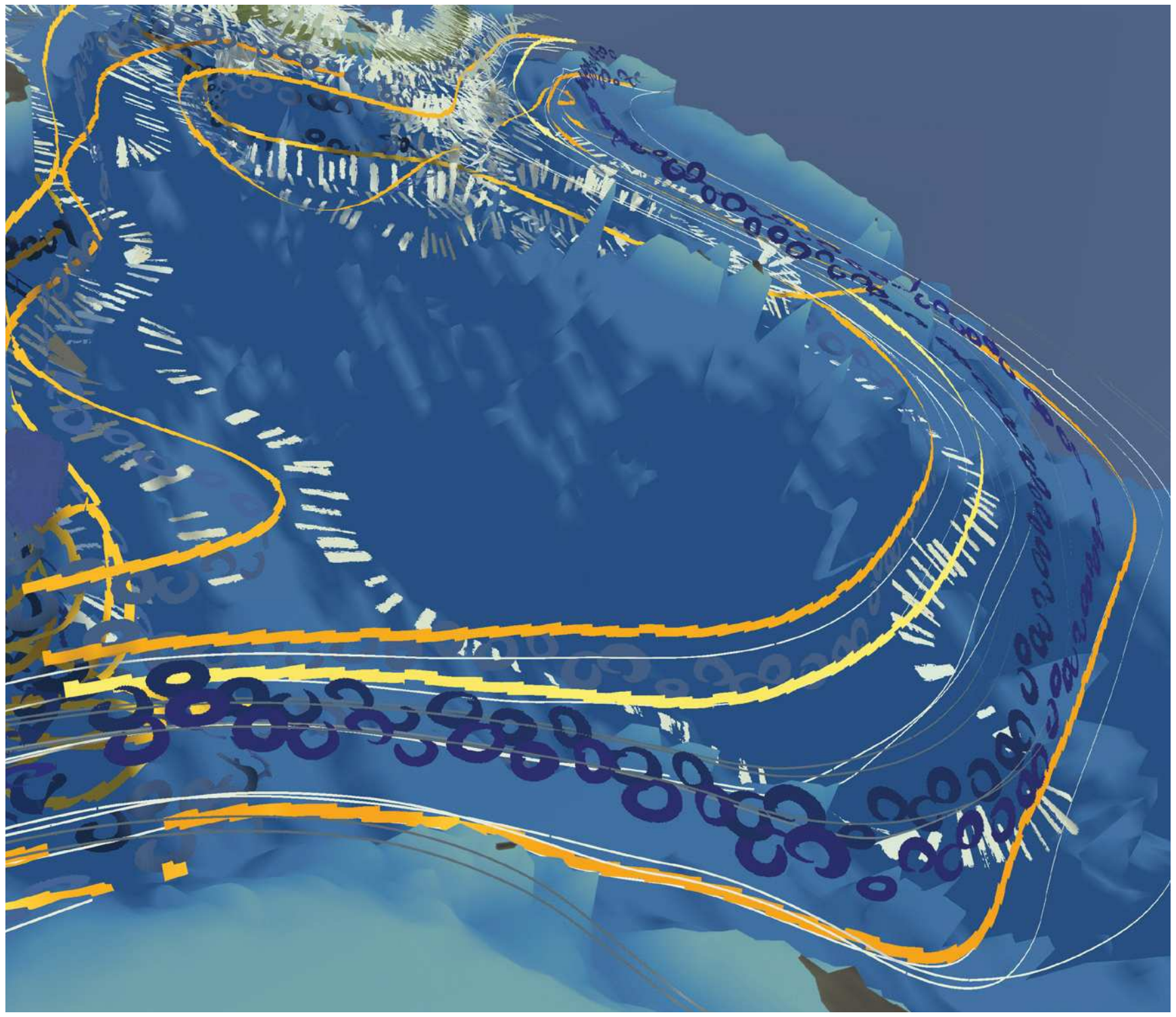

Figure 18: Five categories of line are shown with, colormaps applied to encode scalar data variables that change along the lengths of the lines. The five distinct categories are distinguishable because the lines combine several types of contrast (e.g., linear–curvilinear,

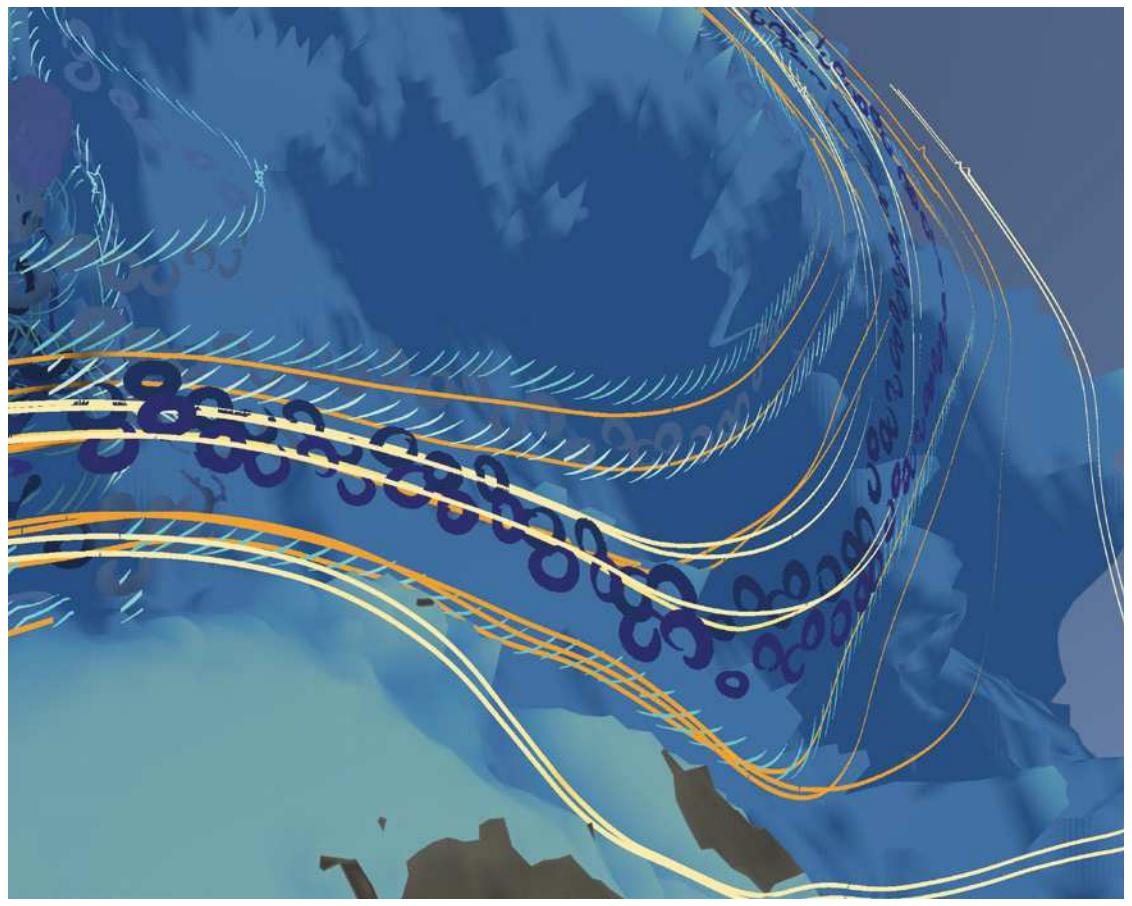

Figure 19: Here is a detail of Figure 15 (right). Here color is applied to the lines. Note that this lowers value contrast between the background and the lines impedes clarity.

## Study of Attention in Hierarchies

In Figure 20, the color and line characteristics combine to create a three-level visual hierarchy. The dark blue line dominates because the blue contrasts in both value and saturation with the background and because complexity of it’s profile contrasts with the simple profiles of the other lines. The turquoise line draws our attention next. The line yellow segmented line, while more complex than the turquoise segments, blends into the background due the similarities in color value and visual texture.

In Figure 21, the prominence of the yellow line is encased by increasing its saturation. Now, we can see the broad coverage of these lines. The line that was originally turquoise now blends with the blue line as they are now both implied lines created with a similar curvilinear, organic mark with more similar colors.

In Figure 22, the width of the yellow line has been increased; it now dominates the composition. While it is, technically, no wider then the other lines, it is solid and continuous creating more weight. The changes can be used to emphasize properties such as the length and strength of these currents.

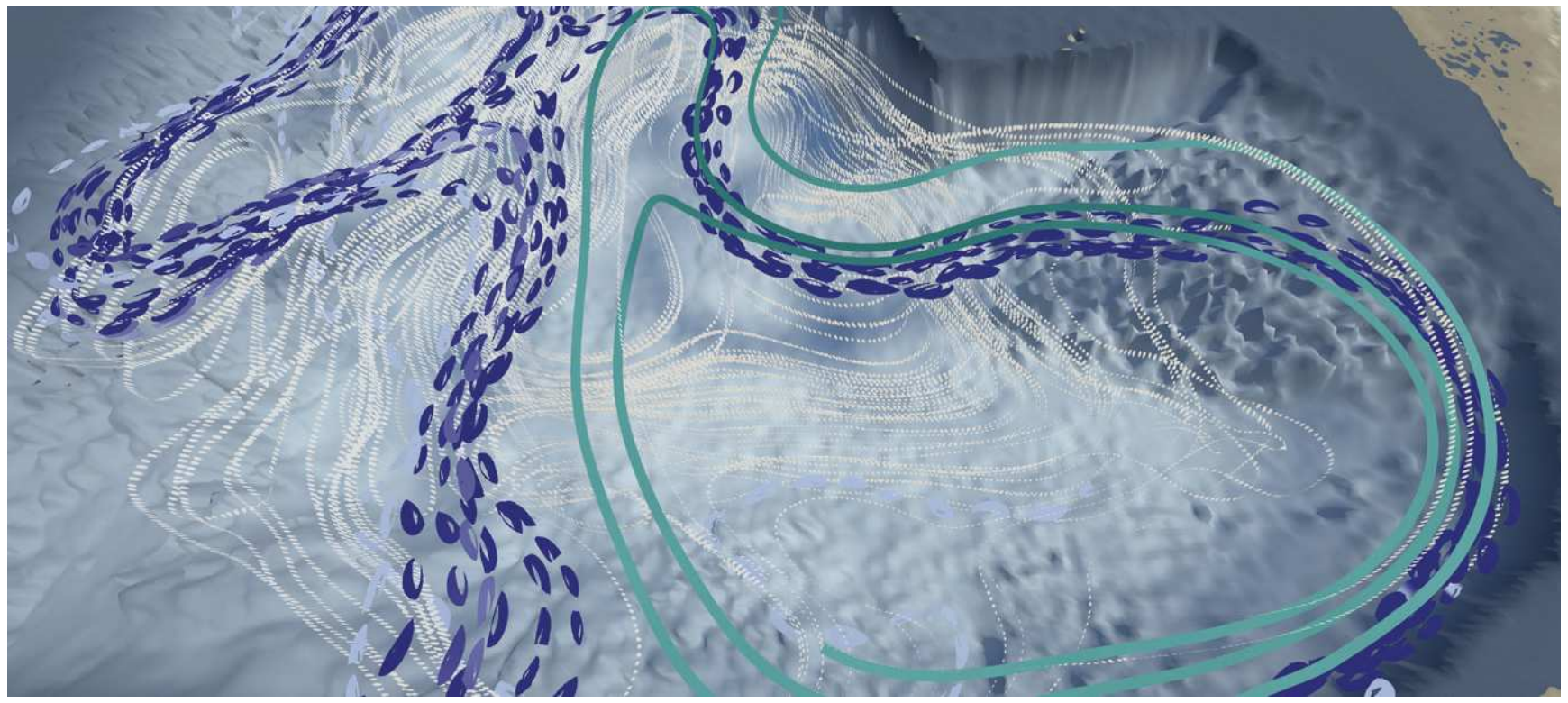

Figure 20: Creating a hierarchy of attention to the three currents shown here.

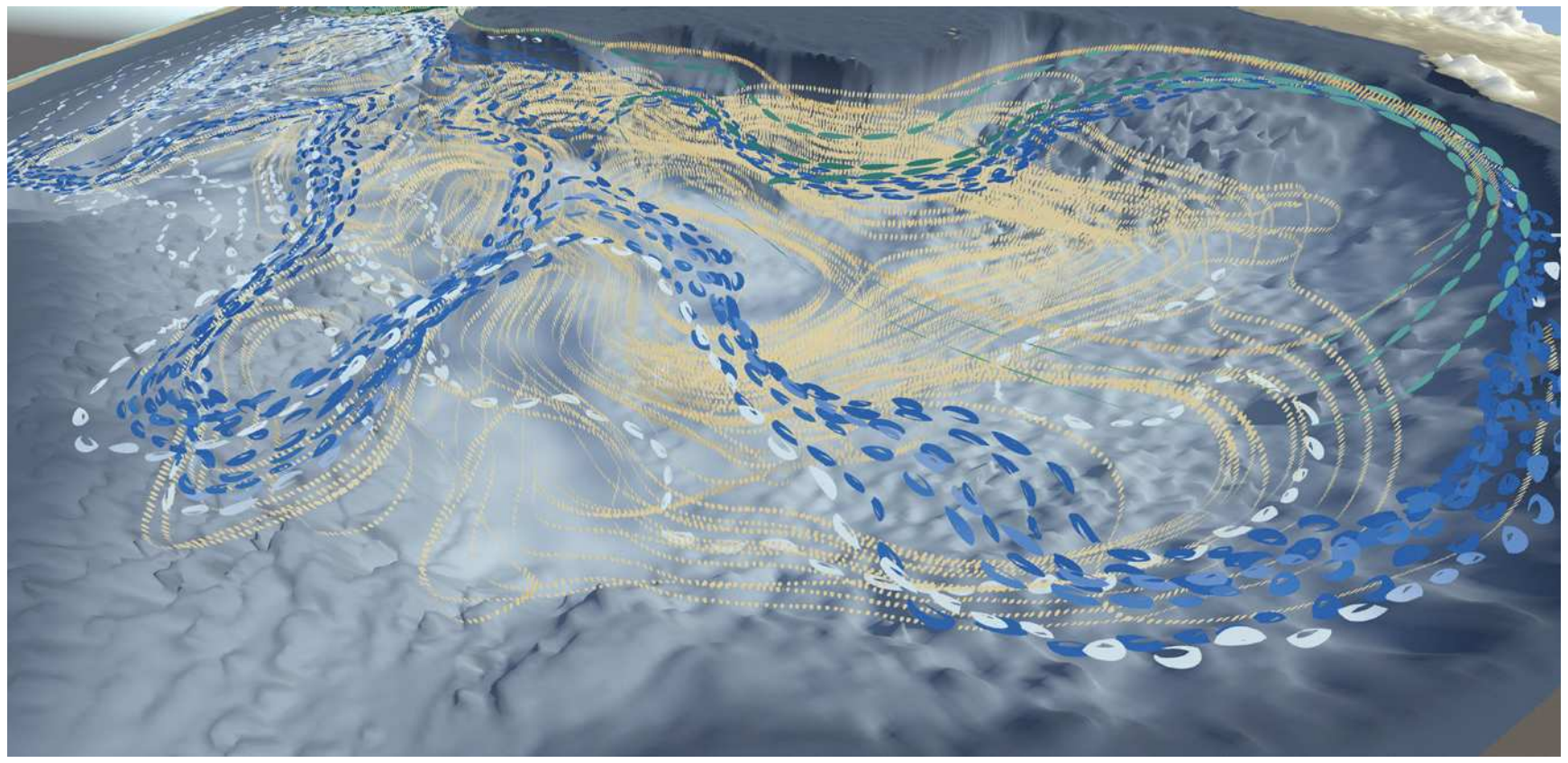

Figure 21: Changes in color intensity.

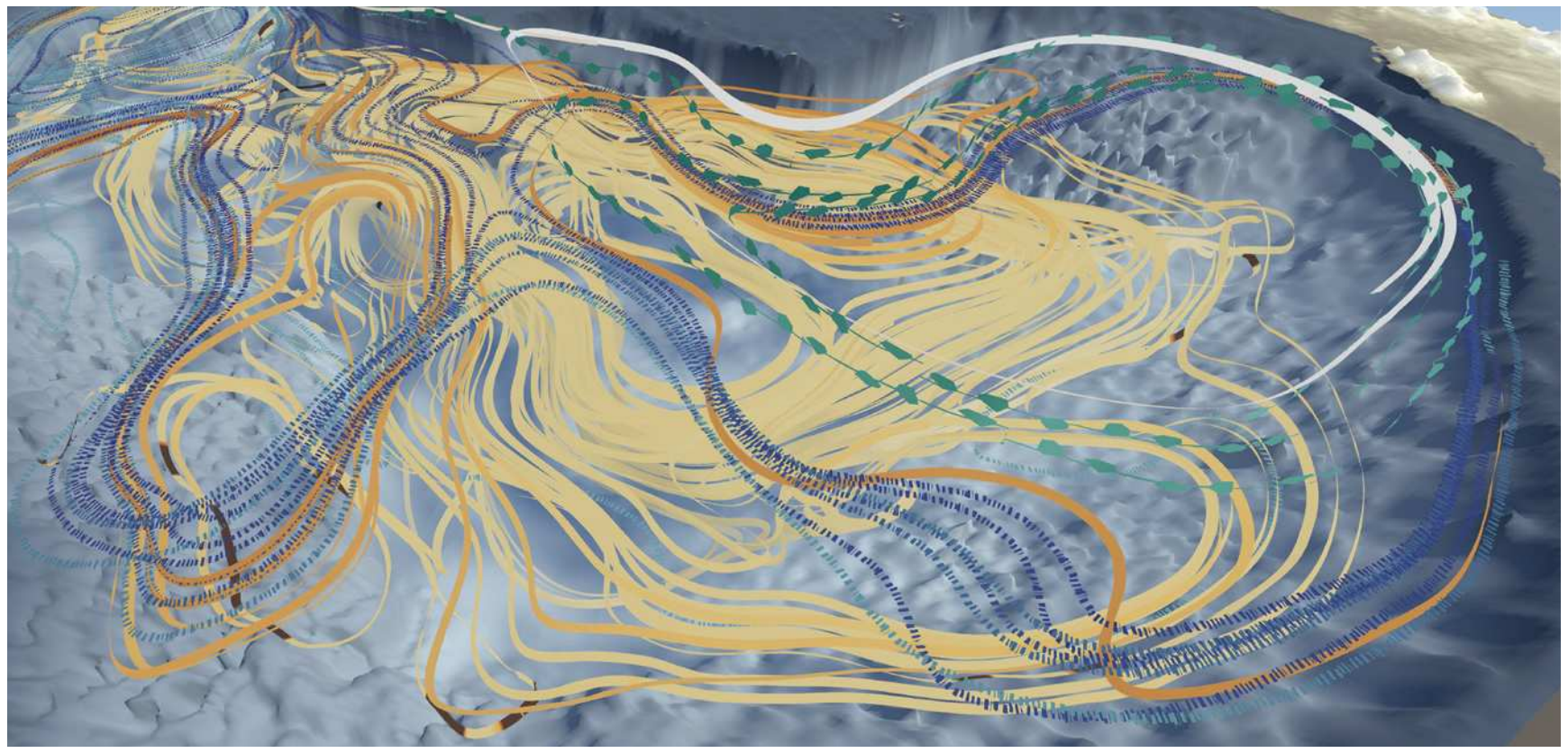

Figure 22: Here attention is focused on the low elevation currents.

## Study of Movement and Direction

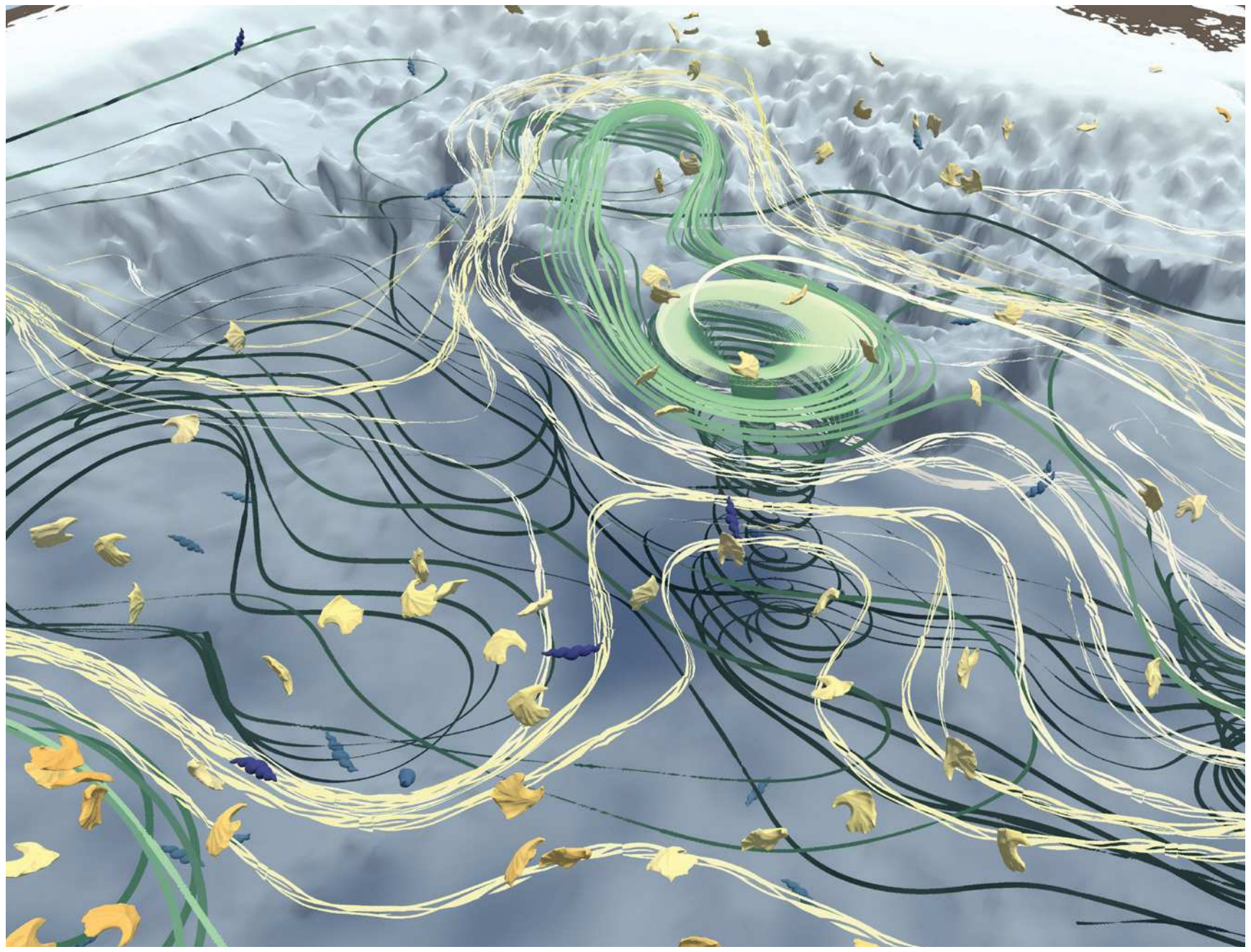

Figure 23: A standard geometric line clearly shows the 3D form of the eddy, but does not convey as much movement as the complex organic lines that imply water current movement. The two styles work well in combination to achieve both purposes.

Figure 24: Experiments with a wider, loose, organic implied line that implies movement. The wide spacing between the marks of the line counterbalance the width, preventing it from overwhelming the other elements of the visualization.

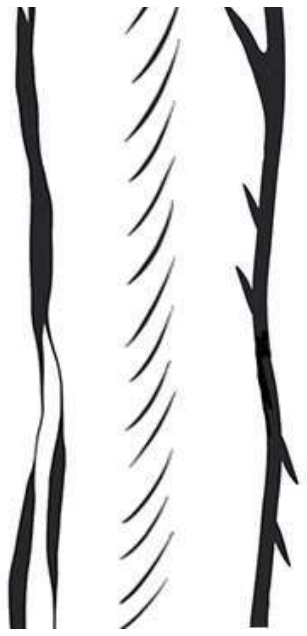

Figure 25: The three line styles used in this study of Movement and Direction.

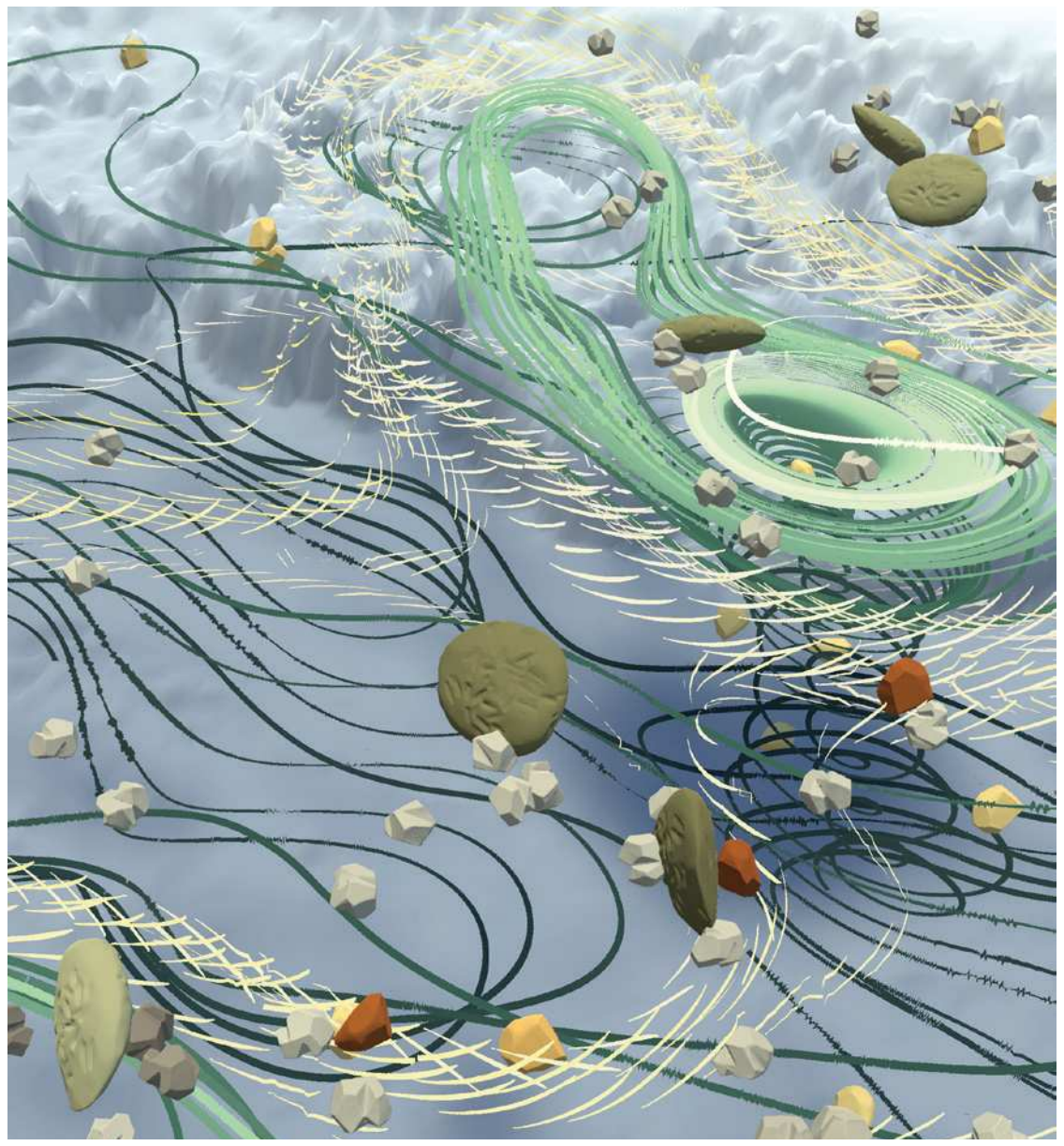

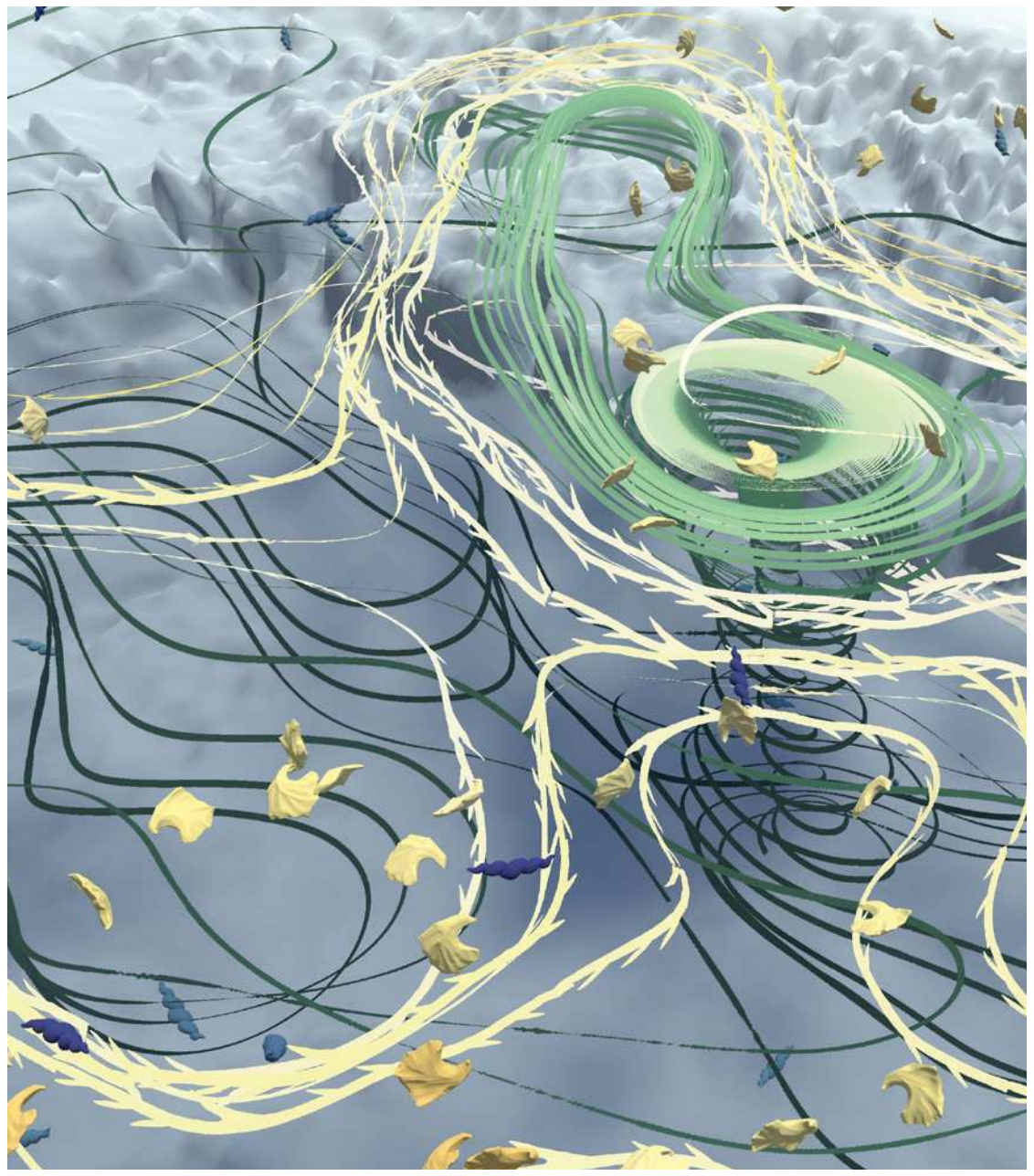

Figure 26: The barbs on the line indicate direction of the flow, which is often an important goal of such visualizations.

Visualizing dense and/or overlapping 3D data often requires a creative approach (e.g., interactivity, animation, and/or thoughtful visual design) to handle occlusion. The implied line constructed with open forms provides a potentially valuable tool for addressing this challenge in that the open regions of the line provide a window through which we can see background data that would otherwise be occluded.

One possible application is to use line to suggest 3D surfaces or other volumetric forms. Curvature-directed surface textures with large open regions have been shown to support more accurate perception of 3D forms when compared to semi-transparent renderings [https://dl.acm.org/doi/abs/10.1145/258734.258796], and the open lines in Figure 24 produce a similar effect. Notice the implied undulating white form in the lower right of the image. These are streamlines calculated from a volumetric vector field; imagine how strong the sense of form would be if they were calculated based upon the principal direction of curvature on a surface.

## Study of Harmony

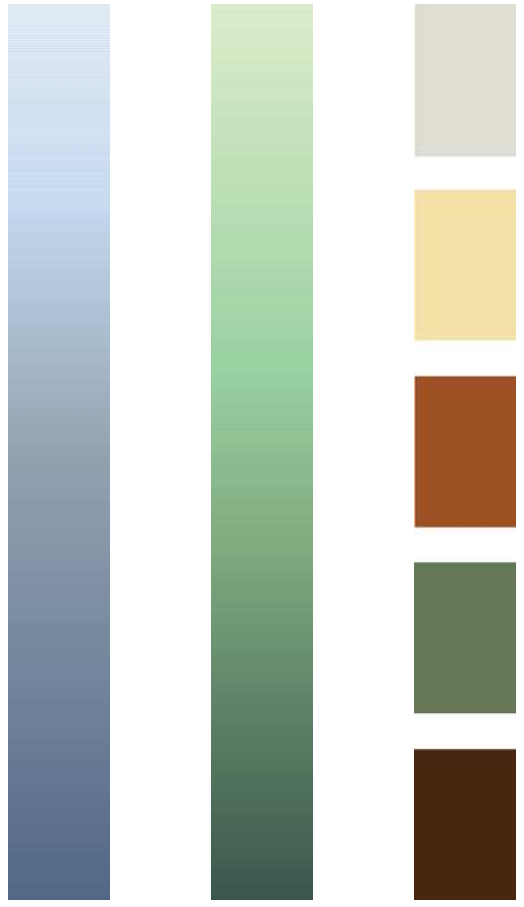

Figure 27: It starts with a quiet, harmonious palette, color intensity and interaction being key to harmony.

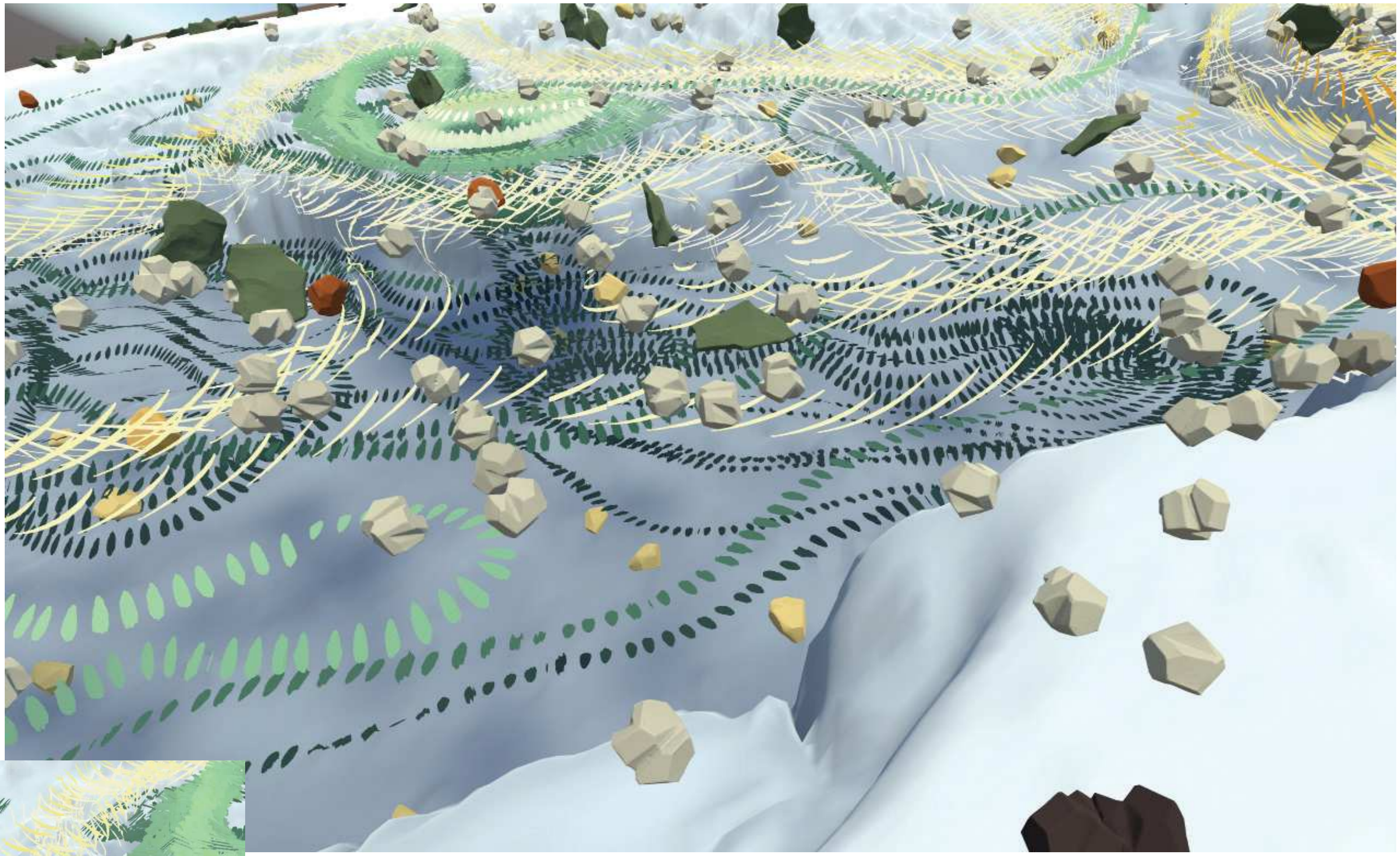

Figure 28: A harmonious composition is created by close attention to the color palette, and to a wide contrast between all of the elements and overcomes the complexity of multiple glyph and line variables.

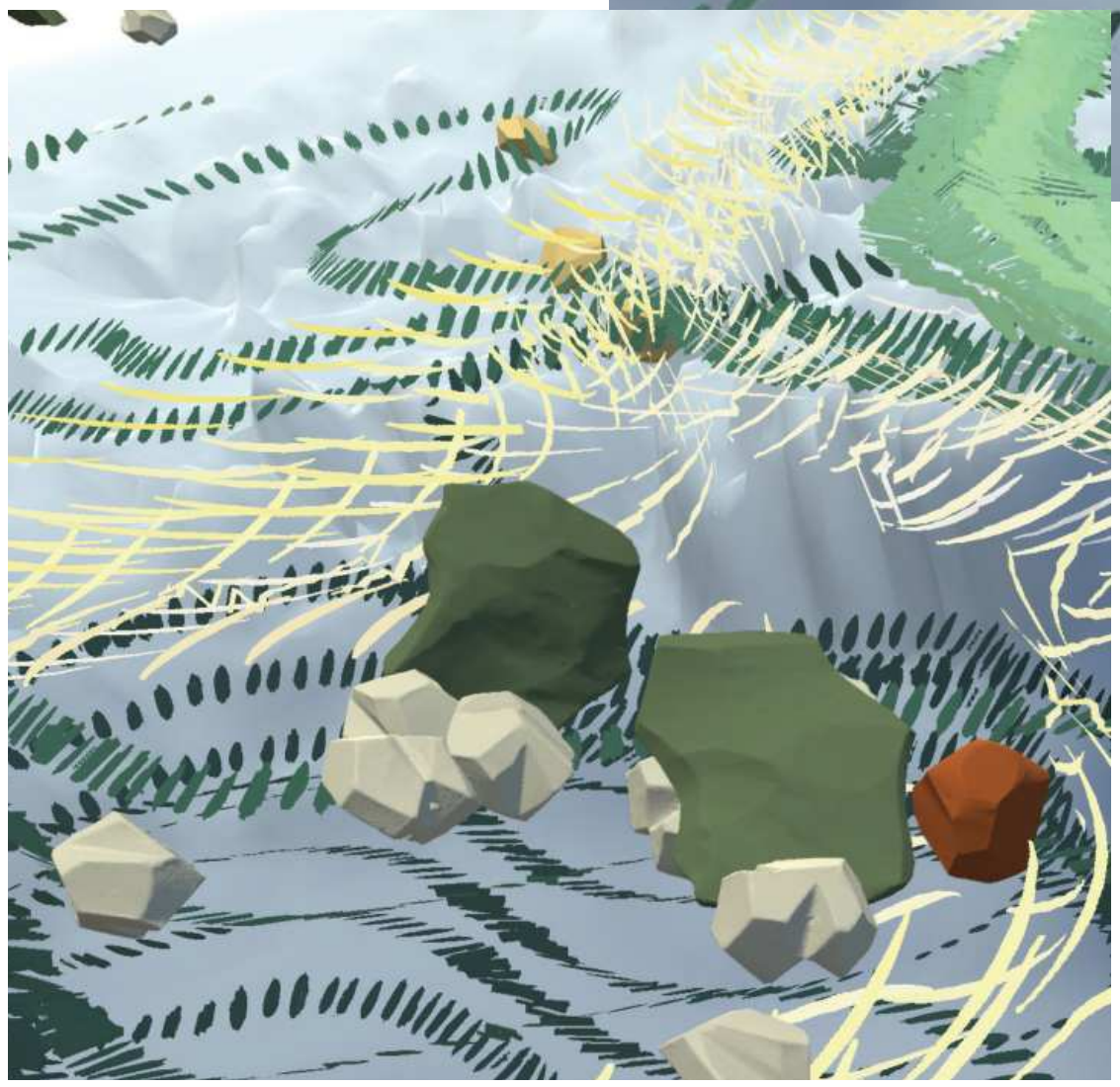

Figure 29: detail from figure 28.

Typically, we would avoid using wide, implied lines in a composition that also includes 3D glyphs because there is a danger that the individual marks that form the implied lines could be confused with the glyphs. However, this visualization is intriguing, and it works because of the strong composition. There are open areas (e.g., lower right) that contrast with the dense areas surrounding the eddy and the small components scattered in the background. These main features are accompanied by layered details that attract and hold our attention: the smaller dark green eddy on the right, the structure of the topography, the way the currents nestle in behind the ridge, the shapes of the glyphs, the analogous palette, and the horizon in the distance.

## Study of Contrast Axes

Each line in this visualization has a distinct function. The chart in Figure 32 outlines the contrast types. Line A is continuous, bold, and aggressive to highlight the path of an important flow. Line B is a light, implied, angled, curvilinear line and functions in an opposite way–it indicates general location and implies movement and currents without dominating. Line C has similar characteristics to Line B but is narrower, more dense and linear. It is less dominant than Line A but easily recognizable and indicates form due to its implied characteristics. Line D operates similarly to Line A; it is a main player in the composition, secondary only to Line A. Line E rests in the background revealing the presence of another set ofcurrents, but it is a set that we must search for, indicating minimal importance.

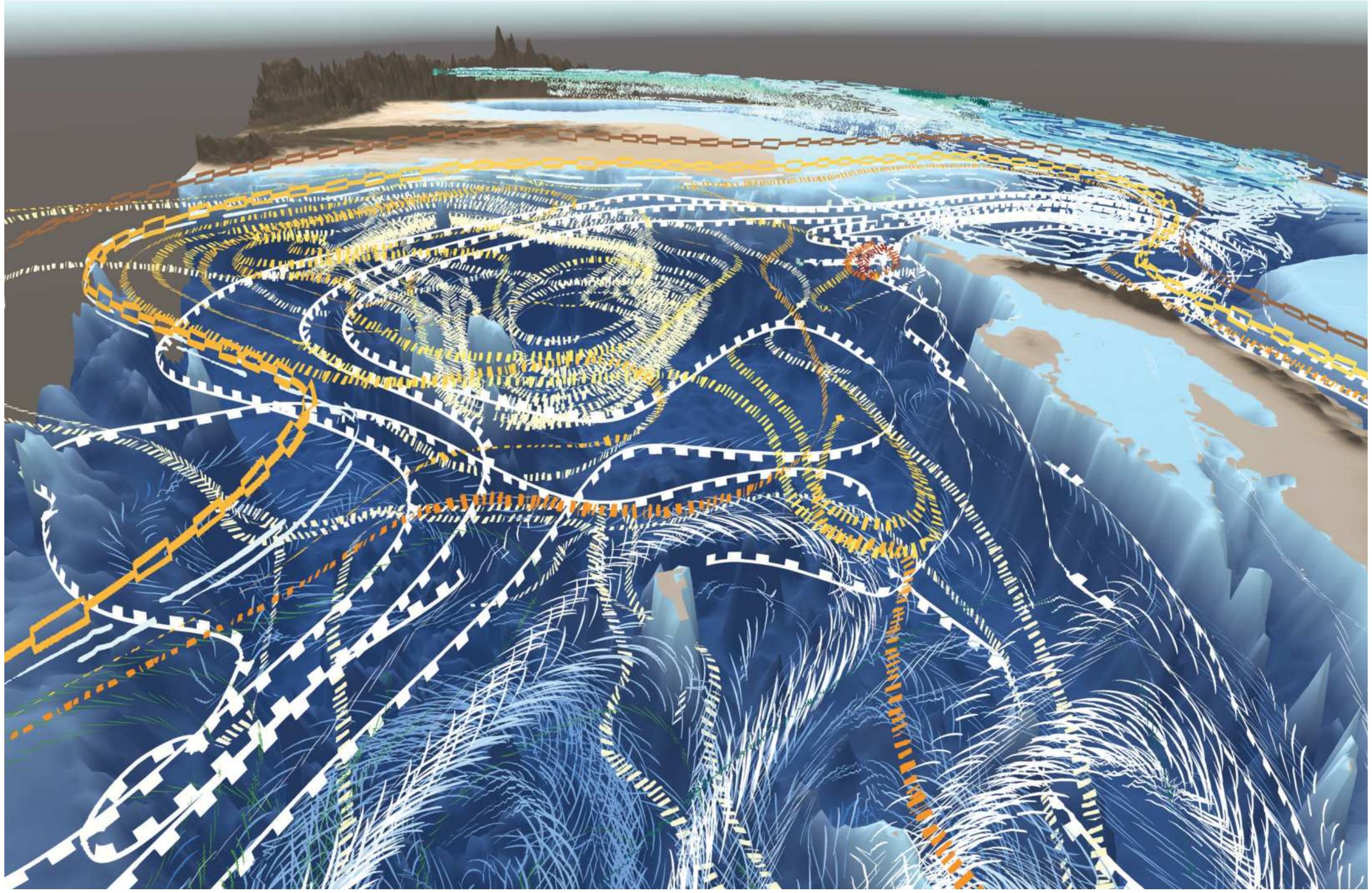

Figure 30: Five lines following the 3D currents, wrapping around the topography, at different elevations.

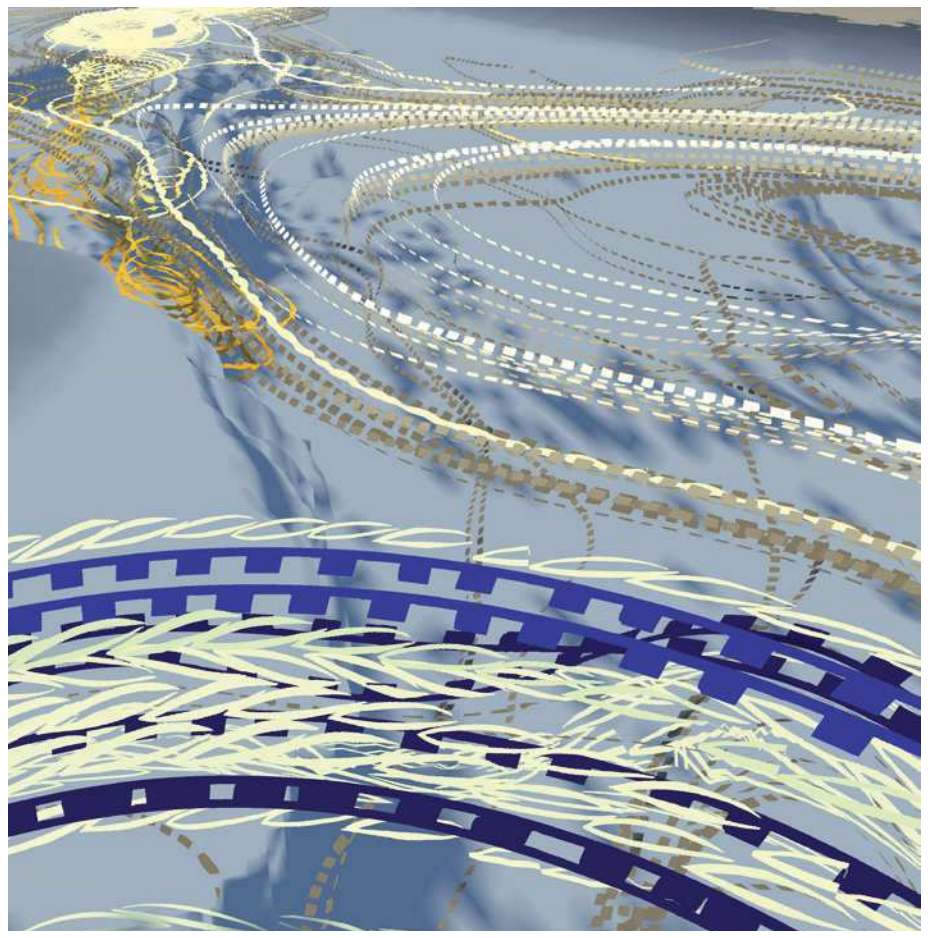

Figure 31: Layered lines differentiated.

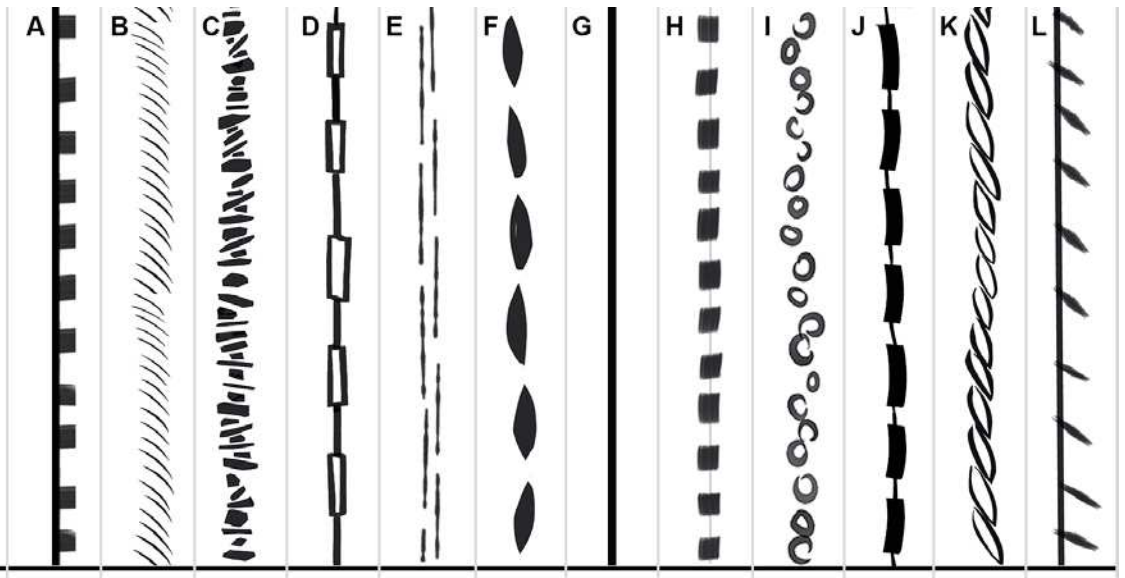

| Line Contrast Categories | A | B | C | D | E | F | G | H | I | J | K | L |
|---|---|---|---|---|---|---|---|---|---|---|---|---|
| Linear(L) - Curvilinear(Cu) | L | Cu | L | L | L | Cu | L | L | Cu | Cu | Cu | L/Cu |
| Open (O) Closed(Cl) | Cl | O | Cl | O | Cl | Cl | Cl | Cl | O | Cl | Cl | O |
| Organic(Or) - Geometric(G) | G | Or | Or | Or | G/Or | Or | G | G/Or | Or | Or | Or | G/Or |
| Continuous(Co) - Implied(I) | Co | I | I | Co | I | I | Co | Co/I | Co/I | Co | Co | Co |
| Thin(T) - Bold(B) | B | T | B | B | T | B | T/B | B | T/B | B | T | T |
| Simple(S) - Complex(Cx) | Cx | Cx | Cx | Cx | S | S | S | S | Cx | Cx | Cx | Cx |

Figure 32: Chart identifying the contrast characteristic of the lines in this section.

Looking back at the contrast categories laid out on page four in Figures 5 and 6, the images in this section illustrate the differentiating line by multiple contrast types. As the number of lines increase so must the contrast ranges. The chart in Figure 30 identifies the contrast category for each of the lines used in this section.

## Study on Allocating Contrast on Multiple Axes

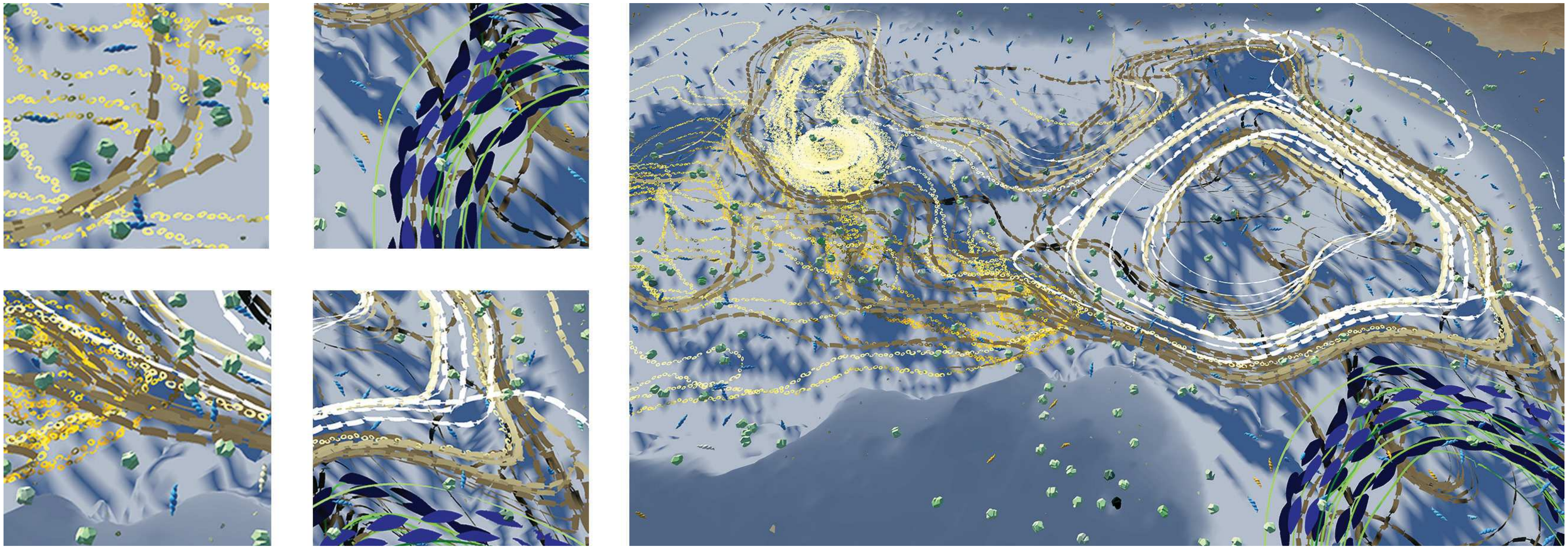

Figure 33: On the left are details from the visualization on the right, where multiple categories of contrast generate clear visual distinction. Color and glyph selection were also driven by contrast theory.

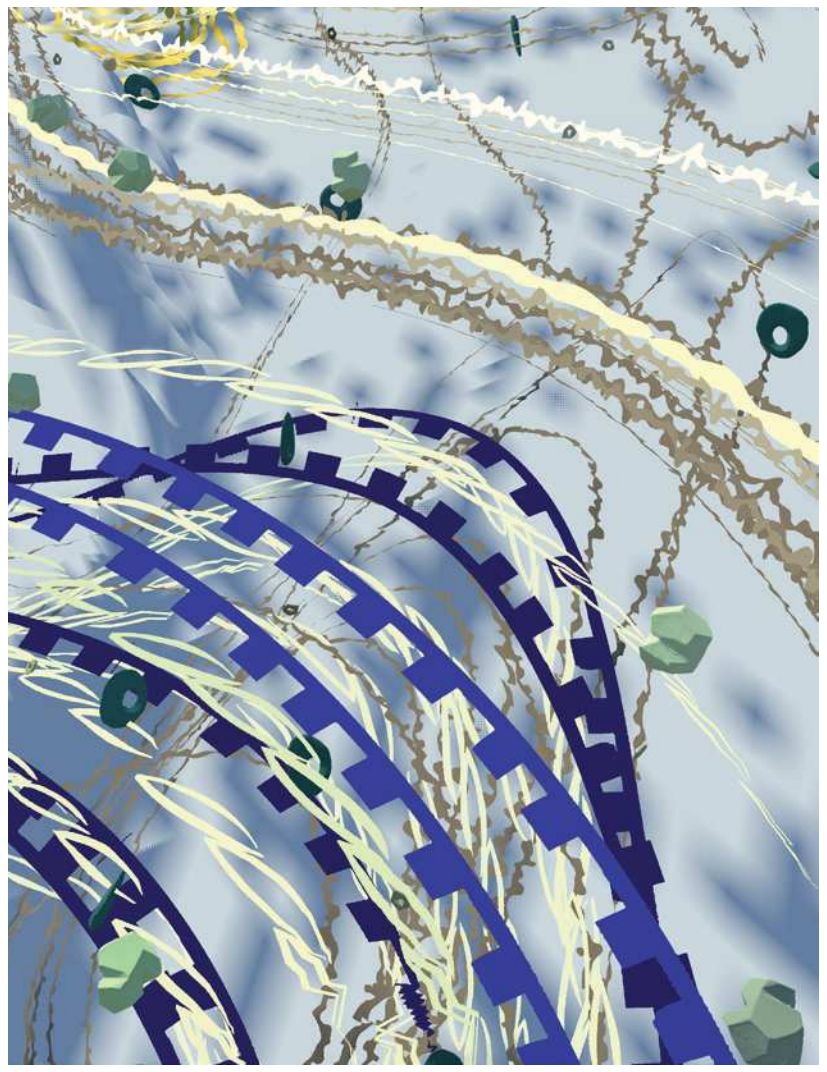

Figure 34: open angular line contrasting with bold saturated blue lines.

In the visualization in Figure 33 there are two areas of overlapping lines. The details on the left illustrate the contrast types detailed in Figure 32 as applied to data.

Figure 34 - a different set of contrasts which includes affect aggressive (blue) and calm (yellow and tan lines), color saturation and glyph selection.

Figure 35 compares an effective contrast with a conflicting contrast example.

Figure 35: The two lines in both of these images each contain multiple characteristics but the image on the right is clearer because the shapes used to construct the lines are similar.

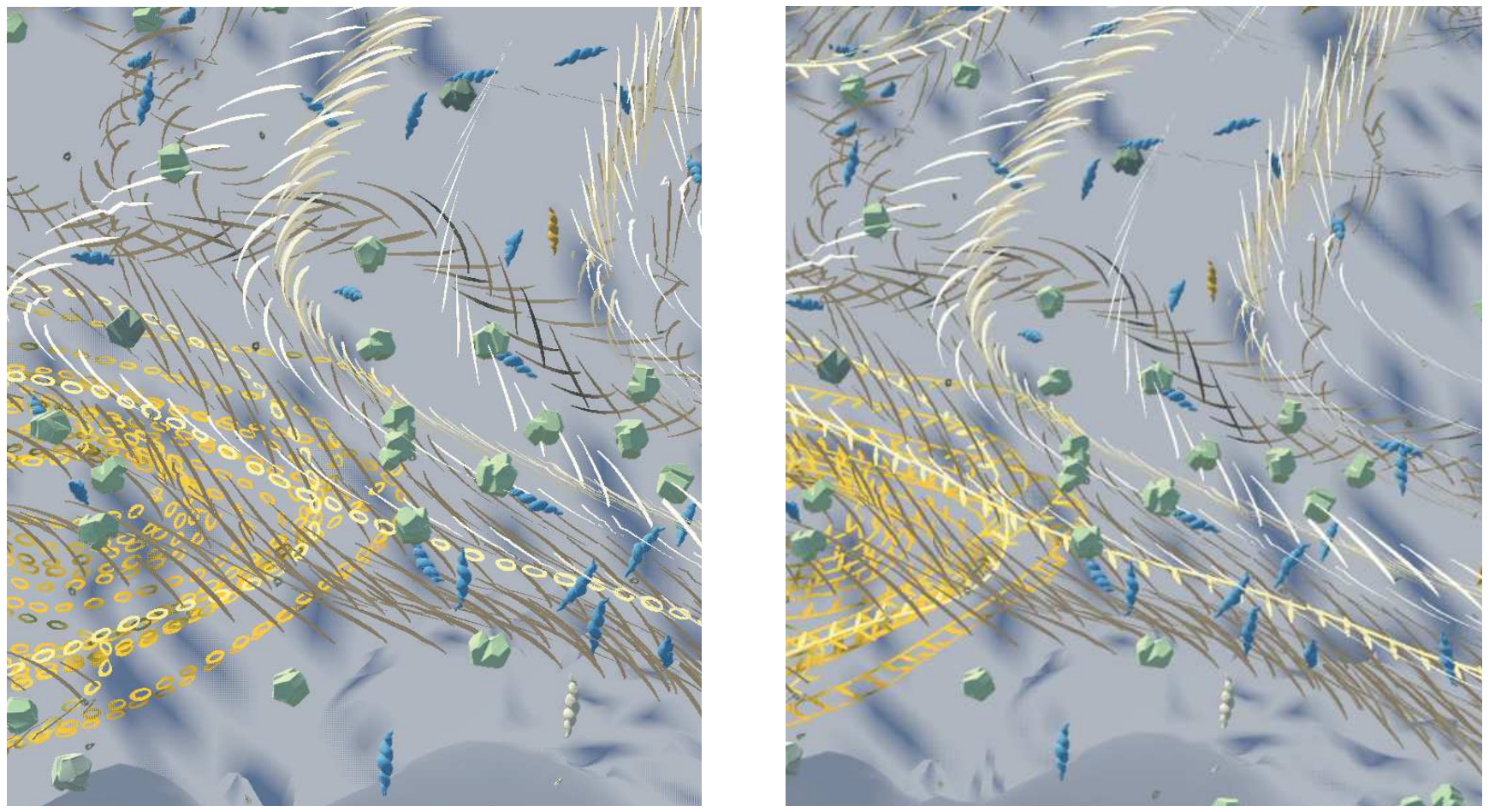

## Study of Implied Association and Uncertainty

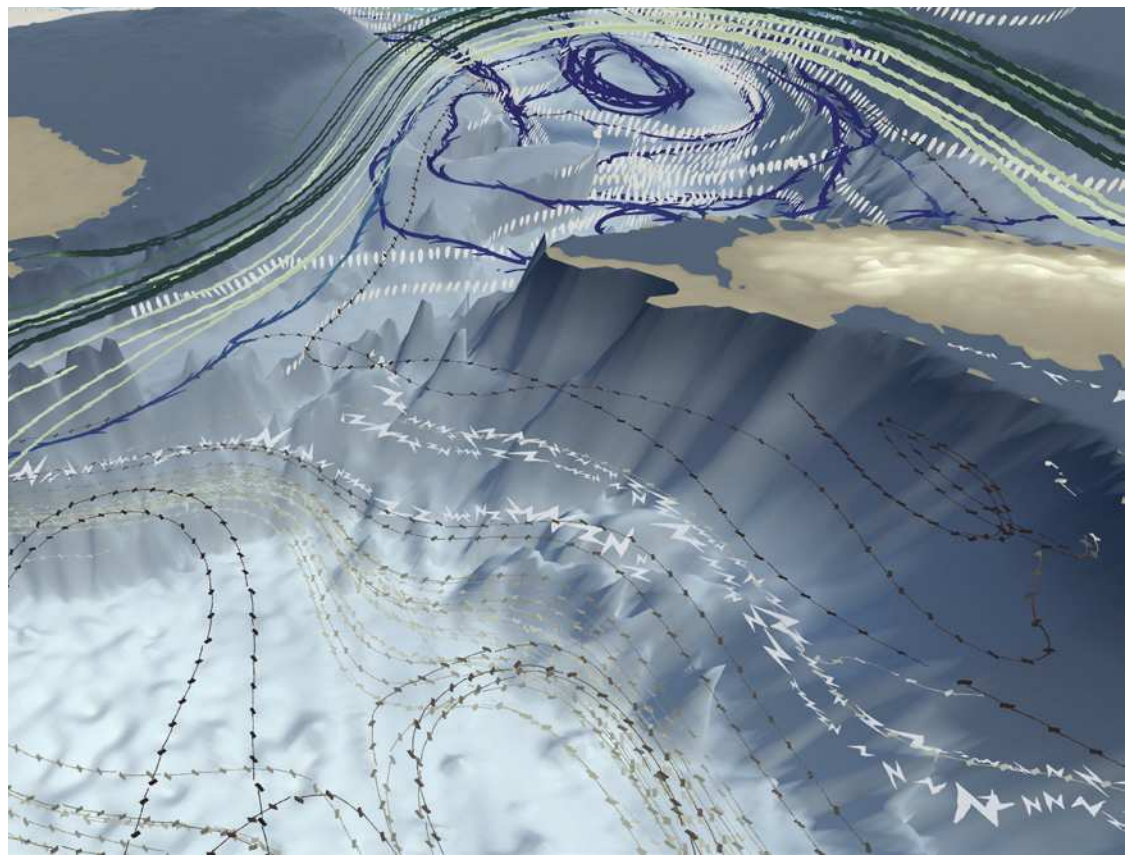

Figure 36: Many visual encodings for data require regularly referencing the legend. However if the encoding reflects the variable's identity or characteristic, that challenge might be removed. For example, these implied lines composed of white electric sparks suggest danger.

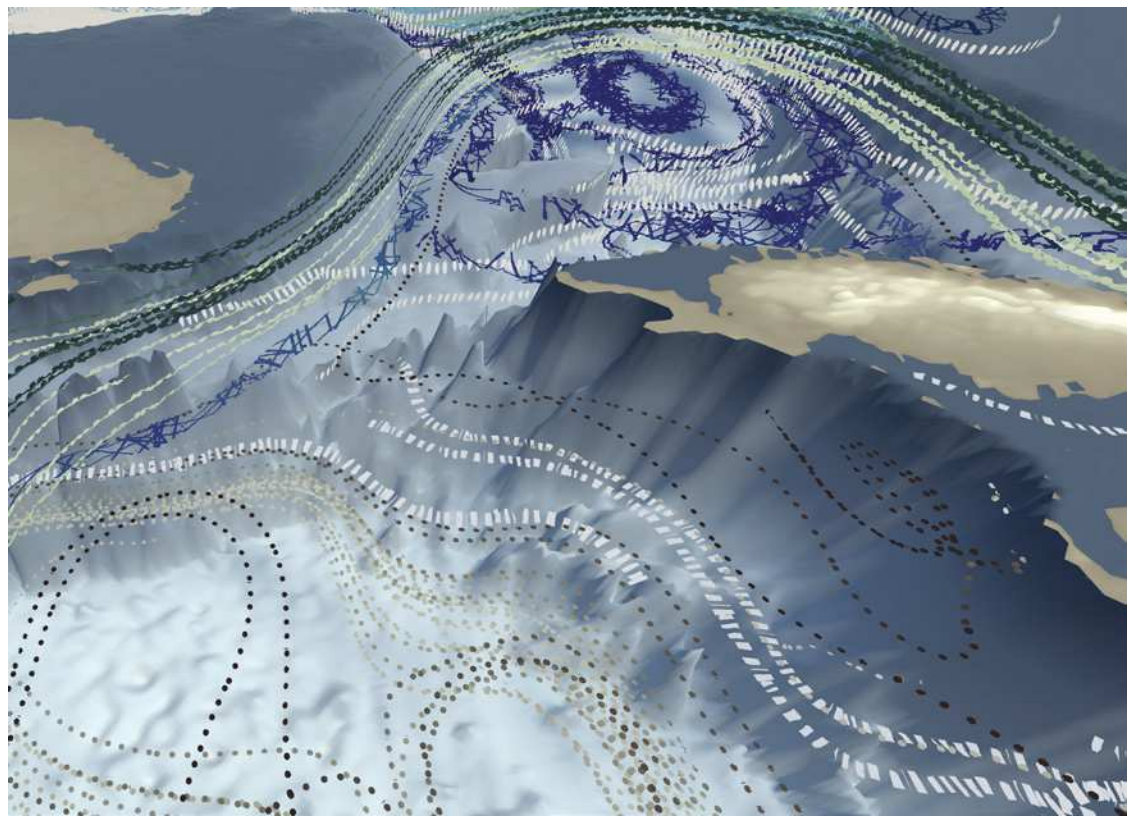

Figure 37: The blue open, linear implied line suggests uncertainty.

Many visualizations require regularly referencing the legend. However if the encoding reflects the variable's identity or characteristic, that challenge can be mitigated as shown in Figure 36 where the electrical symbols imply danger.

Figure 37 demonstrates the opportunity to explore line types able to encode uncertainty.

Figure 38 shows one current encoded with two different line textures. This enables layering a second variable or creating groups of variables. For example, a subset of the currents could be grouped by the presence of a solid line background then differentiated by the texture placed on top.

Figure 39 the complex black line is computer generated. In prior testing [4] this line was considered in a category of its own with associations including messages from outer space, and hieroglyphics. This provides the opportunity to imply separation from the other currents as well as association between them..

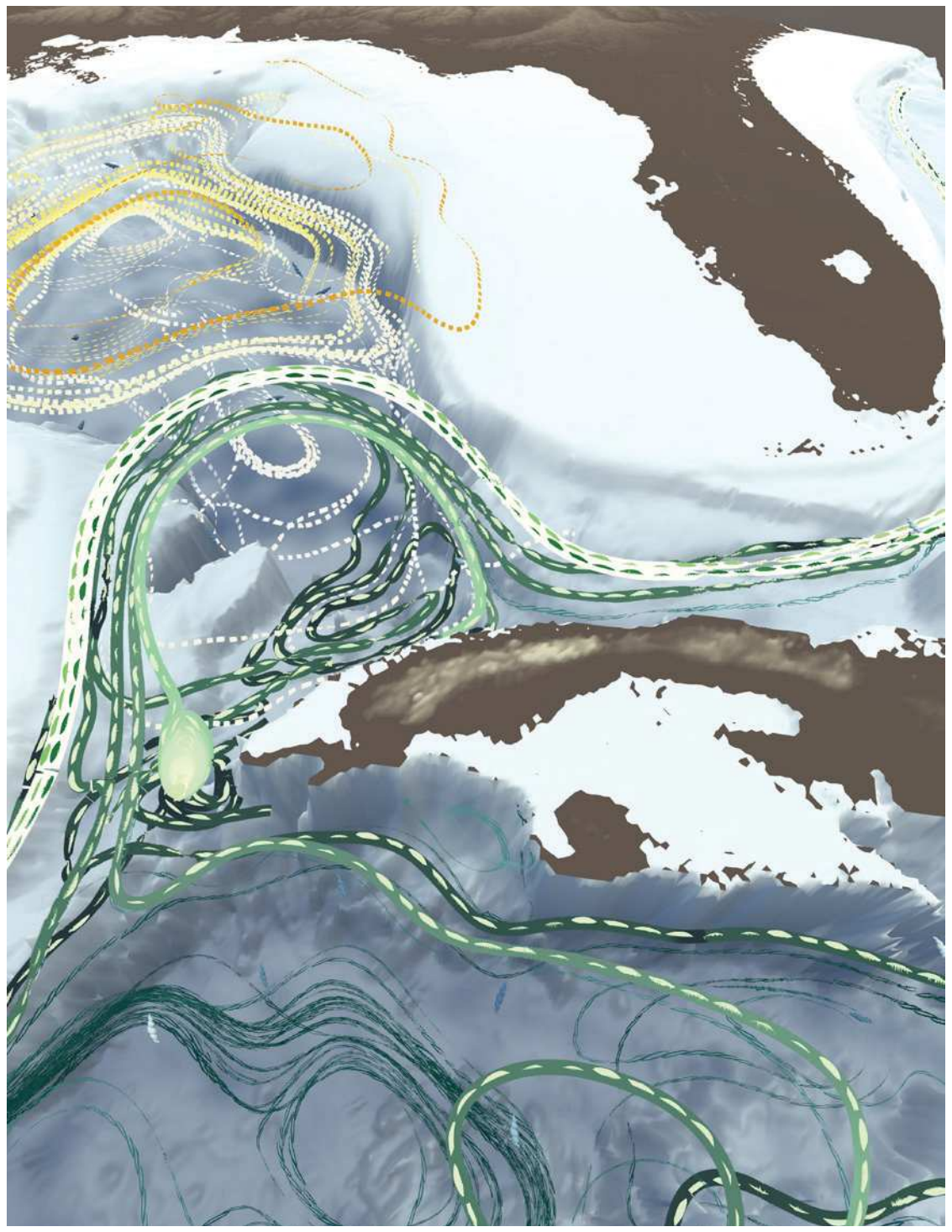

Figure 38: The strongest water current is identified with two overlapping line textures. This opens the opportunity for encoding another dimension of data or categorization. For example, a subset of the currents could be grouped by the presence of a solid line background and then differentiated by the texture placed on top.

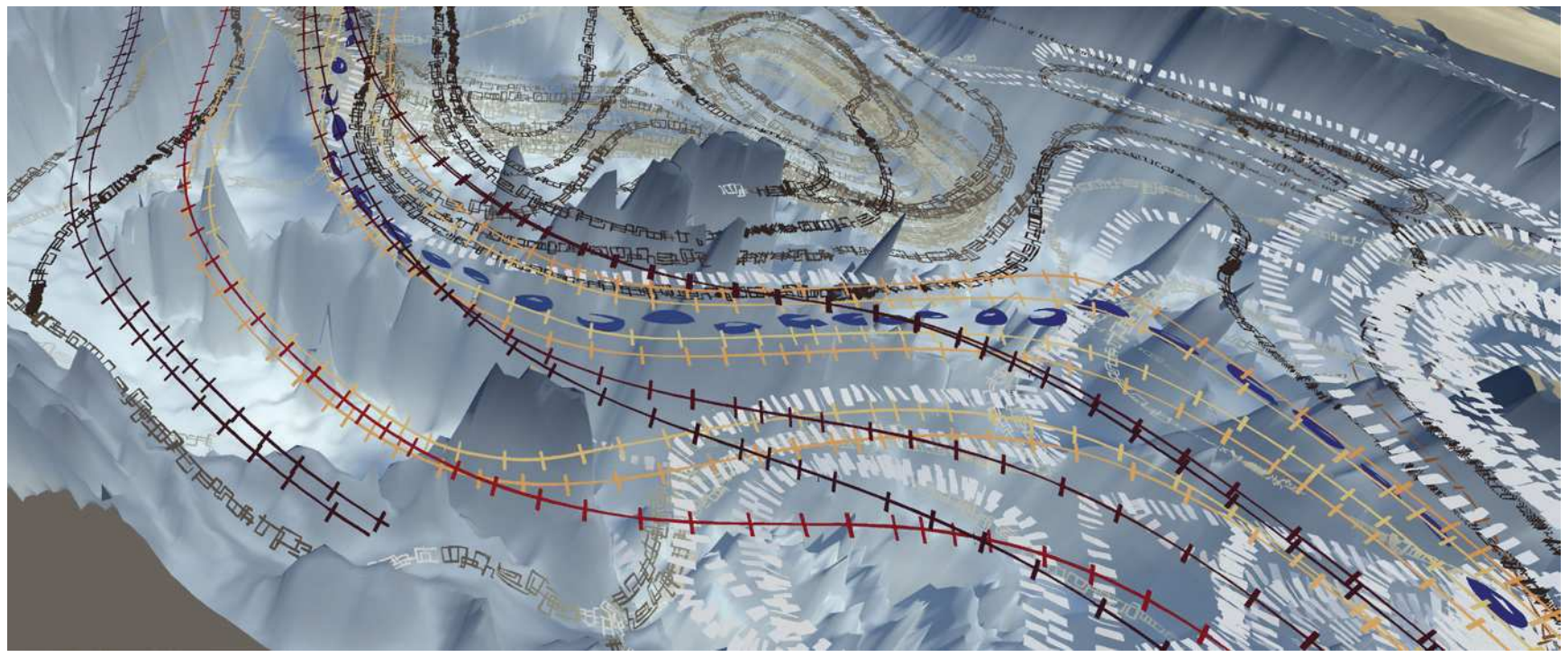

Figure 39: The line incorporating font-like symbols might be associated with hieroglyphics or even messages from outer space.

## Study of Simplicity to Complexity and the Creation of Affect

Figures 40 and 42 explore a different direction. Here, the data and design are relatively simple–just two types of line are used for the visualization. However, in their simplicity, their affect is compelling.

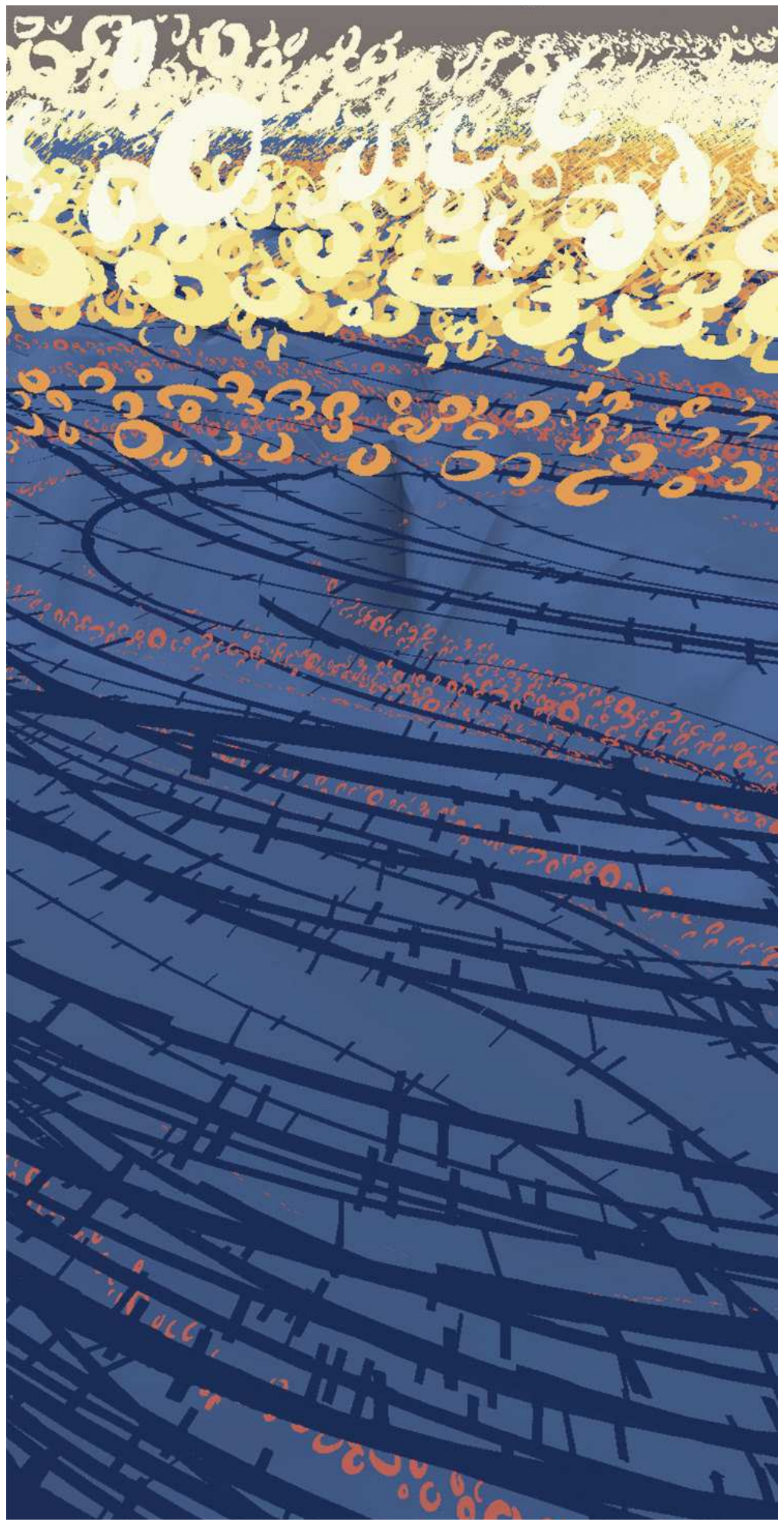

Figure 40: The circles, a gesture drawn line, introduce a narrative.

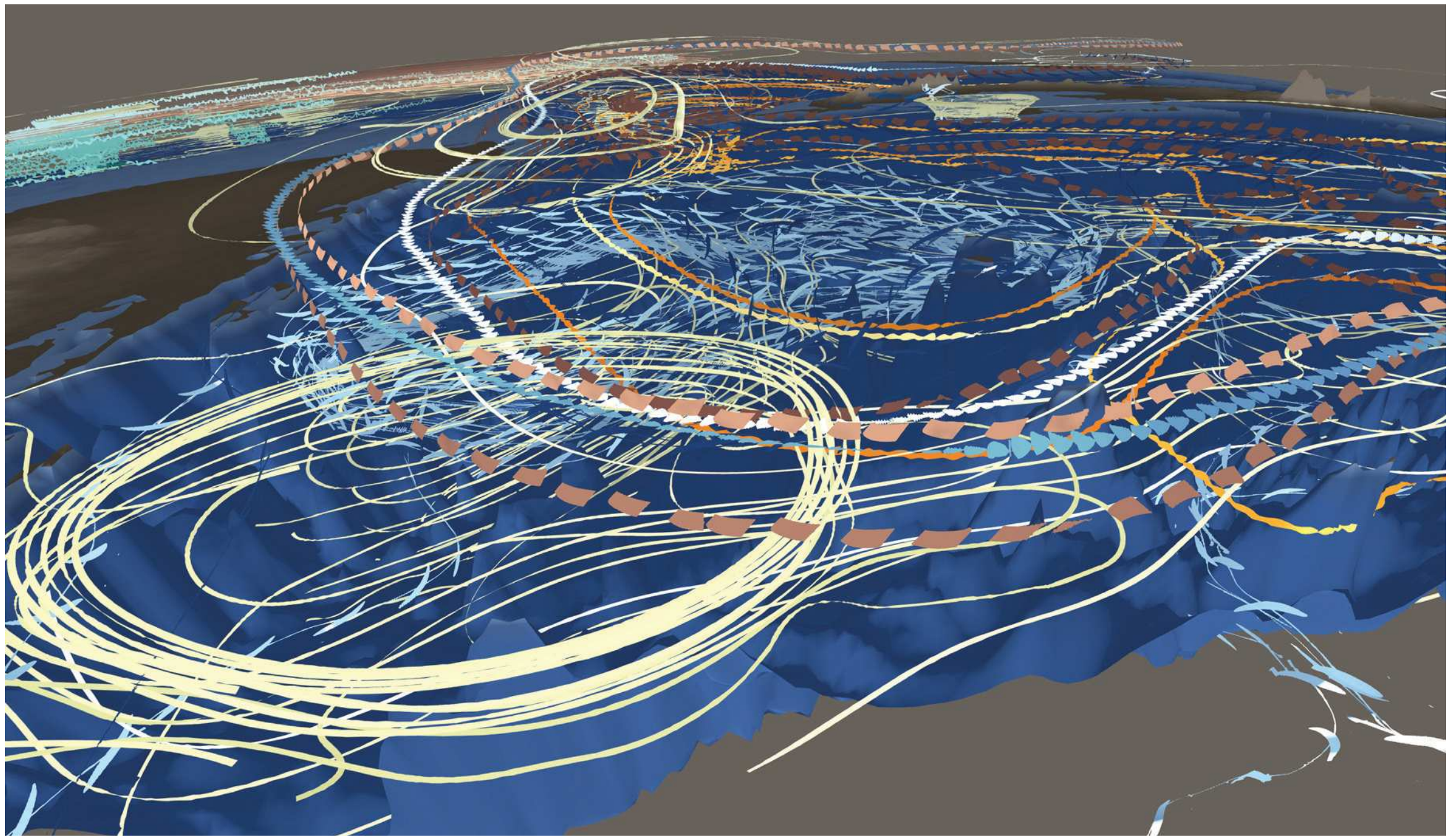

Figure 41: The circles and gestural marks imply a frenzy of intersecting currents and turbulence.

In Figure 42, notice the impact the careful choice of two contrasting line textures can have as you react to the whimsical ovals against the calligraphic, but aggressive, barbed wire, with a hint at the bottom of a broader world that lies just "off the page".

Figure 40 applies the same concepts but to obtain a different affect. Whereas Figures 40 and 42 create a sense of friction and tension, perhaps even suggesting a protagonist and antagonist, Figure 41 is whimsical, like a carefree dance.

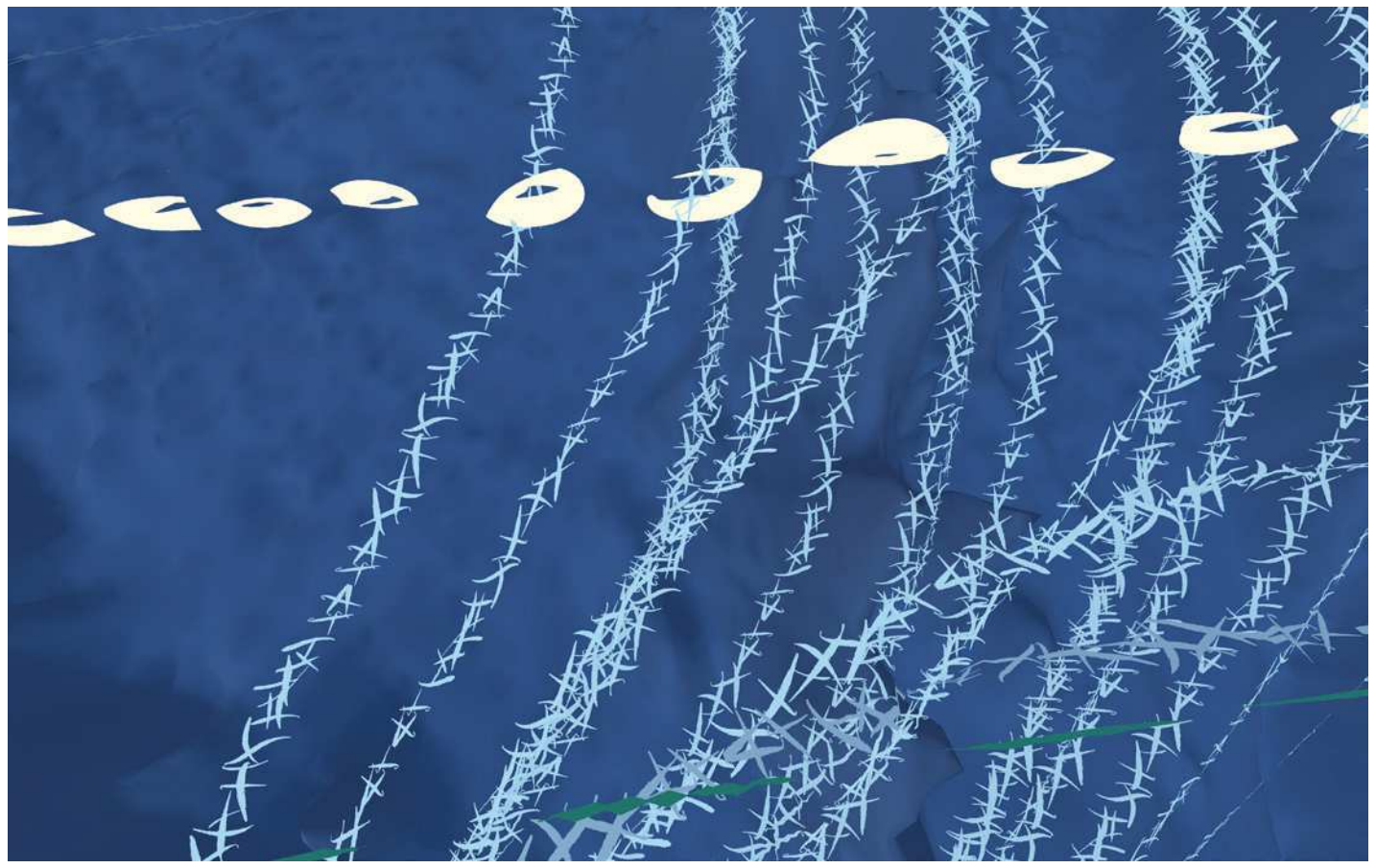

Figure 42 carries an internal power that comes from a combination of line style, color, and composition.

## Conclusion

Fundamentally, this work investigates a more sophisticated visual language for encoding data, and how artists and design theories can guide its use. We believe multivariate volumetric data, like the climate simulations pictured here, are some of the most challenging from the standpoint of visual design. Compared to non-spatial data, where the choice of where visual marks are positioned in space can be powerful, perhaps the most powerful, way to encode data. With spatial data, the designer has no control of the positioning of lines and, thus, must turn to other visual properties to encode data. This powerfully motivates the search for more expressive qualities of line.

We believe the visual results documented here provide strong evidence that sophisticated qualities of line can be effectively controlled to encode data in 3D visualizations. Even subtle changes in line contrasts (e.g., open versus closed, curved versus linear) help differentiate the lines. We found compound contrasts useful and easy to create (e.g., open and linear versus closed and curvilinear).

In general, we found that thin lines work better in dense visualizations. If the lines are too thick, they tend to interfere with one another. There are exceptions, and we include several examples of open, wide lines that functioned well in our compositions, (see Figures 28 and 29). We found it was important to assign colors to create a value (brightness) contrast between the background and the lines and, when possible, to simplify the background to minimize its value range as in Figures 28 and 29. When the background falls into a mid-value range, we found it useful to create lines that are both lower and higher in value as in Figures 14 and 15. We are just beginning to collect user feedback to understand how stakeholders react to visualizations like these [4], and our future work includes both additional design and evaluation.

## Acknowledgements

We would like to thank the entire Sculpting Vis Collaborative team for making this work possible. www.sculpting-vis.org.

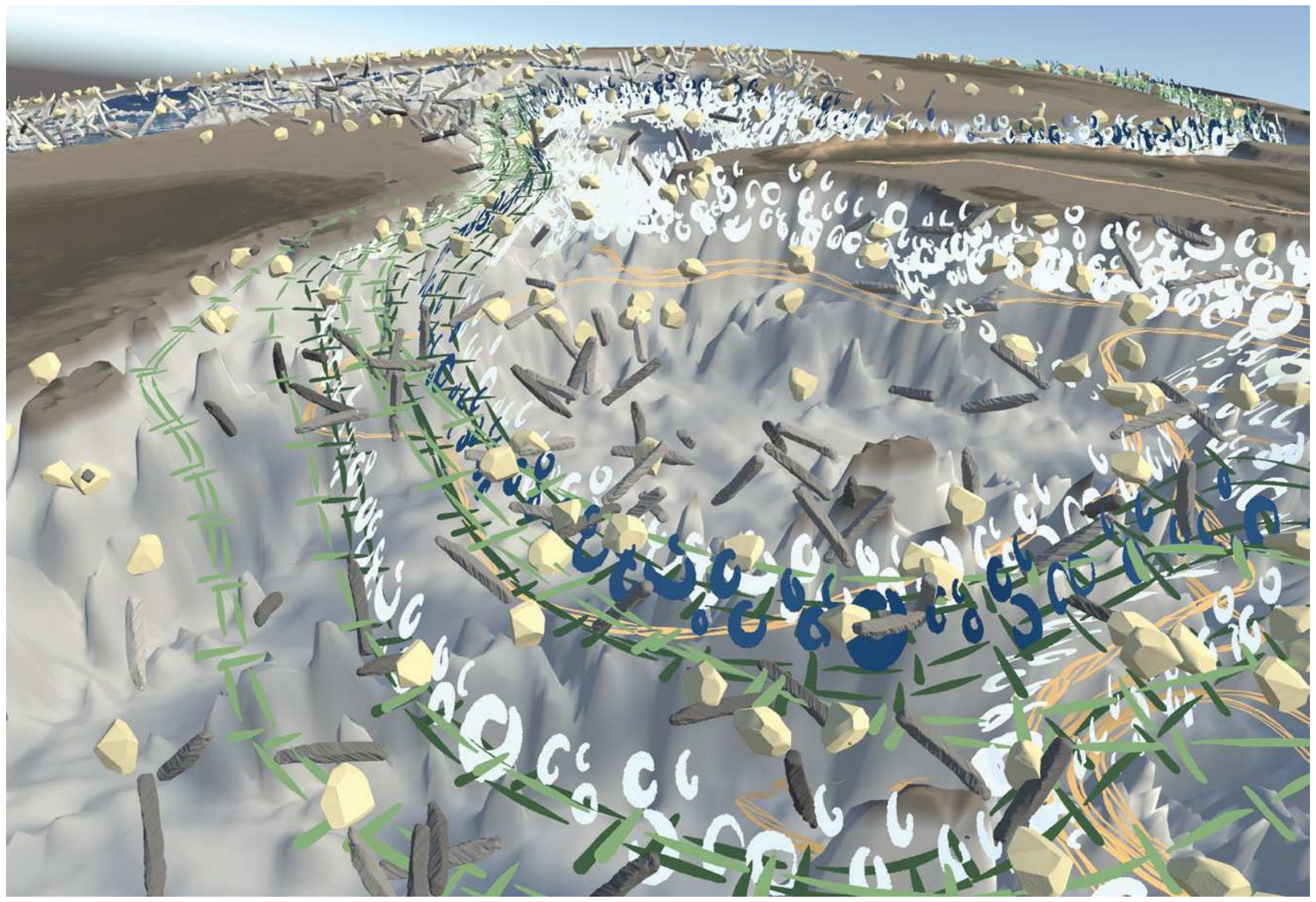

Figure 43: Wrangling the chaos of data toward creating intrigue and clarity in a visual language accessible to us all.